\documentclass{aastex6}
\newcommand{\vlsr}{$v_{\rm LSR}$ }
\newcommand{\kms}{km s$^{-1}$ }

\begin{document}

\title{Linear Polarization of Class I Methanol Masers\\
in Massive Star Forming Regions}

\author{Ji-hyun Kang\altaffilmark{1}, Do-Young Byun, Kee-Tae Kim, Jongsoo Kim\altaffilmark{2}, and Aran Lyo}
\affil{Korea Astronomy and Space Science Institute, 776 Daedeokdae-ro, Yuseong-gu, Daejeon 34055, Republic of Korea}

\and

\author{W. H. T. Vlemmings}
\affil{Department of Earth and Space Sciences, Chalmers University of Technology, Onsala Space Observatory, SE-439 92 Onsala, Sweden}

\altaffiltext{1}{jkang@kasi.re.kr}
\altaffiltext{2}{Korea University of Science and Technology, 217 Gajeong-ro, Yuseong-gu, Daejeon 34113, Republic of Korea}

\begin{abstract}
Class I methanol masers are found to be good tracers of the
interaction between outflows
from massive young stellar objects (YSOs) with their surrounding media.
Although polarization observations of Class II methanol masers
have been able to provide information about
magnetic fields close to the
central (proto)stars, polarization observations of Class I methanol
masers are rare, especially at 44 and 95~GHz.
We present the results of linear polarization observations of 39 Class I
methanol maser sources at 44 and 95 GHz.
These two lines are observed simultaneously
with one of the 21-m Korean VLBI Network (KVN) telescopes in single dish mode.
Approximately 60\% of the observed sources have fractional polarizations
of a few percent in at least one transition.
This is the first reported detection of linear polarization of the 44~GHz methanol maser.
The two maser transitions show similar polarization properties,
indicating that they trace similar magnetic environments,
although the fraction of the linear polarization is slightly higher at 95~GHz.
We discuss the association between
the directions of polarization angles and outflows. We also discuss some
targets having different polarization properties at both lines,
including DR21(OH) and G82.58+0.20, which show the $90^\circ$ polarization
angle flip at 44~GHz.

\end{abstract}


\section{Introduction}
Even though magnetic fields are thought to play an important role
in regulating the formation of stars, there are still many
questions regarding their exact shape, strength, and effect
\citep[e.g.,][]{Crutcher:2012aa}.
For example, statistical research on the field alignment in different scales
in star forming cores,
i.e., from a parsec-scale cloud down to a $\sim 1000$~AU scale envelope,
has recently been performed with
limited number of samples 
and the correlation between the field orientation
and the outflow direction is still under debate
\citep{Chapman:2013aa, Hull:2014aa, Hull:2013aa, Zhang:2014aa, Surcis:2013aa}. 
To reveal the field properties in star forming regions at various 
scales and densities, observations of dust continuum polarization have been used
to study the envelopes
at several thousands AU scales~\citep{Tang:2013aa, Koch:2010aa, Hull:2014aa, Stephens:2014aa}.
On the other hand, masers (OH, H$_2$O, SiO, and CH$_3$OH)
are the only tracers to study the physical conditions
of dense regions $(n_H = 10^{5-11} {\rm cm^{-3}})$
close to protostellar disks or outflows, that are embedded in thick dust envelopes.
For a long time, masers
were thought to trace the isolated pockets of compressed field
rather than the field of an ambient medium. However,
some recent studies of 1.6~GHz hydroxyl and 6.7 GHz methanol masers
showed that they possibly probe
the large scale magnetic field~\citep[e.g.,][]{Vlemmings:2010aa,
Fish:2006aa, Surcis:2012aa}.

Class I methanol masers, including 44 and 95 GHz transitions, 
are known to trace the regions where the outflow
of massive young stellar object (YSO) is interacting
with the surrounding medium, while Class II methanol masers are
tightly correlated with the central (proto)stars
\citep{Plambeck:1990ab, Kurtz:2004aa, Gomez-Ruiz:2016aa, Menten:1991aa}.
Water masers are also related to outflows, but they appear to 
arise from shocked regions closer
to the central object than Class~I methanol masers
\citep[e.g.,][]{Kurtz:2004aa}.
The polarimetric observations of Class II methanol masers have been
increased during the last decade~\citep[e.g.,][]{Vlemmings:2012aa}
and have been able to provide information about the magnetic fields near
the central objects.
However, the polarization studies of Class I methanol masers have been
very limited.
For example, the detection of linear polarizations at 95~GHz
has been reported only for two sources 
~\citep{Wiesemeyer:2004aa}.
No linear polarization has been reported for 44~GHz methanol maser
sources, whereas circular polarization at 44~GHz, detected with the
Very Large Array (VLA), has been presented
only for one object, OMC-2~\citep{Sarma:2011aa}.
Therefore, further linear and circular polarization observations 
of the Class I methanol masers need to be performed.

In this paper, we report the single-dish measurements of 
linear polarization toward 44 and 95~GHz methanol maser sources. 
It is not straightforward to interpret the single dish polarization
results. Many maser features with different polarization properties
within a large single dish beam are
blended in a spectrum. Furthermore, the fractional polarization is
affected by maser saturation as well as the angle between the maser
propagation direction
and the, yet unknown, magnetic field direction along the line-of-sight.
However, single-dish polarization observations still contribute
significantly to improving our understanding of the masers, and the
magnetic field in the maser region. Specifically, polarization results
for large source samples
provide statistical view on the polarization properties of 
44 and 95~GHz methanol masers. 
As the 44 and 95~GHz masers are believed to trace similar regions
\citep[e.g.,][]{Valtts:2000aa, Kang:2015ab},
their polarization angles are expected
to be similar. By observing the two simultaneously,
we can examine whether this is the case. Additionally,
\cite{Wiesemeyer:2004aa} have detected high fraction
of linear polarizations ($> 30$\%) for some of their targets,
which requires specific conditions of anisotropic pumping
or loss mechanism for high $J$ transitions, where $J$ is a rotational
quantum number. With further observations,
we can verify whether these conditions are indeed regularly achieved.

This paper is structured as follows.
Section 2 describes source selection criteria, observation and reduction
methods. Sections 3 presents the statistics of polarization results.
We discuss the implication of the measured polarization properties, the similarity or difference of the
polarization properties of 44 and 95~GHz maser, and the possible association of maser polarization
angle with larger scale fields in section 4. We discuss some individual targets with
some unique polarization properties in section 5. We summarize our study in section 6.

\section{Observations and Data Reduction}
\subsection{Target Selection and Coordinate Refinement}
We observed 39 sources and they are listed in Table~\ref{tb1}.
We selected 36 bright Class~I methanol maser sources with peak flux densities
$\gtrsim 50$~Jy from the KVN 44~GHz
methanol maser survey~\citep{Kim:2012aa}.
We also observed three weaker sources (S231, W51Met2, W75S(3)),
for which linear polarizations were
detected by \cite{Wiesemeyer:2004aa} in Class~I methanol maser transitions
but were not included in the KVN 44~GHz methanol maser survey.

\floattable
\begin{deluxetable}{lccrrrrrrrr}
\tabletypesize{\footnotesize}
\tablecaption{Source Summary \label{tb1}}
\tablehead{
\colhead{} & \colhead{} &\colhead{} &
\multicolumn{2}{c}{$F\nu$} &\multicolumn{2}{c}{$v_{\rm LSR}$} &
\multicolumn{2}{c}{$\sigma$} & \colhead{} & \colhead{} \\
\colhead{Source} & \colhead{$\alpha_{2000}$} &
\colhead{$\delta_{2000}$} & \colhead{44} & \colhead{95} &
\colhead{44} & \colhead{95} & \colhead{44} & \colhead{95} & \colhead{Pol} & \colhead{} \\ 
\colhead{Name} & \colhead{( h:m:s )} & \colhead{($^\circ:\arcmin:\arcsec$)} &
\multicolumn{2}{c}{( Jy )} & \multicolumn{2}{c}{( km s$^{-1}$ )} &
\multicolumn{2}{c}{( Jy )} & \colhead{44/95} & \colhead{Ref} 
}
\startdata
 OMC2 &05:35:27.124 & $-$05:09:52.48 & 210 & 290 & $ +11.1$ & $ +11.1$ &
 0.36 &  0.71& Y/Y & (1)\\
 S231\tablenotemark{a} &05:39:13.060 & +35:45:51.30 &  26 &  22 & $ -16.7$ & $ -16.7$ &
 0.25 &  0.72& N/N & (2) \\
 S235 &05:40:53.250 & +35:41:46.90 &  74 &  60 & $ -16.4$ & $ -16.4$ &
 0.31 &  0.83& N/N & (1) \\
 S255N &06:12:53.630 & +18:00:25.10 & 260 & 190 & $ +11.0$ & $ +11.0$ &
 0.30 &  0.52& Y/Y &(1)\\
 NGC2264 &06:41:08.073 & +09:29:39.99 & 170 & 120 & $  +7.2$ & $  +7.2$ &
 0.33 &  0.85&  N/Y &(1)\\
 G357.96-0.16 &17:41:20.140 & $-$30:45:14.40 &  66 &  47 & $  -5.0$ & $
  -5.0$ &  0.42 &  0.69& Y/Y &(3)\\
 G359.61-0.24 &17:45:39.080 & $-$29:23:29.00 &  99 &  53 & $ +19.3$ & $
 +19.3$ &  0.47 &  1.21& Y/N &(1) \\
 G0.67-0.02 &17:47:19.230 & $-$28:22:14.50 &  33 &  30 & $ +69.2$ & $ +65.5
$ &  1.94 &  4.38& N/N &(1)\\
 IRAS18018-2426\tablenotemark{b} &18:04:53.010 & $-$24:26:40.50 & 690 & 260 & $ +10.9$ & $
 +10.8$ &  0.52 &  0.88& Y/Y &(3)\\
 G10.34-0.14 &18:09:00.000 & $-$20:03:35.00 & 110 &  78 & $ +14.6$ & $ +14.6
$ &  0.39 &  0.66& Y/Y &(3)\\
 G10.32-0.26 &18:09:22.860 & $-$20:08:05.80 & 170 &  69 & $ +32.4$ & $ +32.5
$ &  0.50 &  0.86& Y/N &(3) \\
 G10.62-0.38 &18:10:29.070 & $-$19:55:48.20 & 100 &  28 & $  -6.8$ & $  -6.7
$ &  0.60 &  2.14& N/N &(1)\\
 G11.92-0.61 &18:13:58.100 & $-$18:54:31.00 &  67 &  30 & $ +35.2$ & $ +35.2
$ &  0.60 &  1.52& N/N &(4)\\
 W33MET &18:14:11.183 & $-$17:56:00.00 &  54 &  36 & $ +32.8$ & $ +32.8$ &
 0.81 &  1.81&  N/Y &(4)\\
 G13.66-0.60 &18:17:24.080 & $-$17:22:14.10 &  65 &  47 & $ +48.3$ & $ +48.4
$ &  0.73 &  1.88& N/N &(1)\\
 G18.34+1.78SW &18:17:49.950 & $-$12:08:06.48 & 620 & 390 & $ +30.4$ & $
 +30.3$ &  0.43 &  0.77& Y/Y &(3)\\
 G14.33-0.63 &18:18:54.200 & $-$16:48:00.50 & 170 & 100 & $ +23.3$ & $ +23.4
$ &  0.83 &  2.16& N/N &(4)\\
 GGD27 &18:19:12.450 & $-$20:47:24.80 & 130 &  58 & $ +13.2$ & $ +13.1$ &
 0.60 &  1.56& Y/Y &(1) \\
 G19.36-0.03 &18:26:25.800 & $-$12:03:57.00 & 130 &  79 & $ +26.5$ & $ +26.2
$ &  0.44 &  0.84& Y/N &(1) \\
 L379 &18:29:23.307 & $-$15:15:29.92 & 140 &  77 & $ +17.6$ & $ +17.6$ &
 0.51 &  0.92& Y/Y &(4)\\
 G25.65+1.04 &18:34:20.910 & $-$05:59:40.50 &  63 &  60 & $ +41.6$ & $ +41.6
$ &  0.46 &  1.09& N/N &(4)\\
 G23.43-0.18 &18:34:39.270 & $-$08:31:39.00 &  71 &  32 & $+101.3$ & $+102.3
$ &  0.32 &  1.91& Y/Y &(3)\\
 G25.82-0.17 &18:39:03.630 & $-$06:24:09.50 &  83 &  81 & $ +90.4$ & $ +90.2
$ &  0.38 &  2.48& N/N &(1)\\
 G27.36-0.16 &18:41:50.980 & $-$05:01:28.00 & 180 & 100 & $ +94.0$ & $ +94.1
$ &  0.36 &  0.85& Y/Y &(4)\\
 G28.37-0.07MM1 &18:42:52.100 & $-$03:59:45.00 &  61 &  32 & $ +76.4$ & $
 +76.5$ &  0.48 &  0.80& N/N &(3)\\
 G28.39+0.08 &18:42:54.500 & $-$04:00:04.00 &  49 &  29 & $ +79.4$ & $ +79.4
$ &  0.39 &  0.64& N/N &(1)\\
 G29.91-0.03 &18:46:05.370 & $-$02:42:17.10 & 410 & 220 & $ +98.2$ & $ +98.2
$ &  0.42 &  0.77& Y/Y &(4)\\
 G30.82-0.05 &18:47:46.840 & $-$01:54:14.00 &  46 &  32 & $ +96.9$ & $ +96.6
$ &  0.39 &  1.26& Y/N &(4) \\
 G40.25-0.19 &19:05:41.440 & $+$06:26:08.00 & 130 & 120 & $ +72.7$ & $ +72.7
$ &  0.33 &  0.58& Y/Y  &(4)\\
 G49.49-0.39 &19:23:43.960 & $+$14:30:31.00 &  97 &  45 & $ +49.1$ & $ +55.7
$ &  0.88 &  2.31& Y/Y &(1)\\
 W51MET2\tablenotemark{a} &19:23:46.500 & $+$14:29:41.00 &  32 &  25 & $ +56.6$ & $ +56.6$ &
 0.37 &  0.84& N/N &(2)\\
 G59.79+0.63 &19:41:03.100 & $+$24:01:15.00 & 110 &  67 & $ +30.8$ & $ +30.8
$ &  0.22 &  0.67& N/N &(1)\\
 W75N &20:38:37.190 & $+$42:38:08.00 &  25 &  13 & $  +8.8$ & $  +8.9$ &
 0.62 &  0.97& N/N &(4)\\
 DR21W &20:38:54.805 & $+$42:19:22.50 & 230 & 210 & $  -2.6$ & $  -2.5$ &
 0.21 &  0.48& Y/N  &(1)\\
 DR21(OH) &20:38:59.280 & $+$42:22:48.70 & 390 & 310 & $  -0.1$ & $  +0.0
$ &  0.24 &  1.35& Y/Y &(1)\\
 DR21 &20:39:01.760 & $+$42:19:21.10 &  88 &  61 & $  -3.7$ & $  -3.8$ &
 0.37 &  0.76& Y/N  &(1)\\
 W75S(3)\tablenotemark{a} &20:39:03.500 & $+$42:26:00.20 &  50 &  30 & $  -5.1$ & $  -5.2$ &
 0.74 &  1.79& N/N &(4)\\
 G82.58+0.20 &20:43:28.480 & $+$42:50:00.90 & 240 & 170 & $ +10.3$ & $ +10.3
$ &  0.20 &  0.64& Y/Y&(1) \\
 NGC7538 &23:13:42.000 & $+$61:27:29.70 &  57 &  20 & $ -57.4$ & $ -57.3$ &
 0.26 &  0.75& N/N &(4)\\
\enddata
\tablenotetext{a}{It is not included in the KVN 44~GHz methanol maser
survey, but its linear polarization was detected by \citet{Wiesemeyer:2004aa}.}
\tablenotetext{b}{A separate maser source, M8E, is located at $15\arcsec$
from the position of IRAS18018-2426.}
\tablecomments{$F_\nu$ is the peak value in the Stokes I spectrum,
and \vlsr is the LSR velocity of the peak channel. The rms $(\sigma)$
was measured in the noise channels in the Stokes I spectrum.
The Y or N in the Pol column indicates whether the linear polarization
for each source is detected or not. The last column shows the coordinate
references.
(1) is the position of the KVN 44~GHz methanol maser survey \citep{Kim:2012aa}.
(2) is from \citet{Wiesemeyer:2004aa}.
(3) is the position adopted from the KVN fring survey (Kim et al. in prep.).
(4) is the new position determined by the grid mapping in this paper.
} 
\end{deluxetable}

Except for some targets observed with the VLA at 44~GHz
\citep{Kogan:1998aa, Kurtz:2004aa},
many of the coordinates used in the KVN 44~GHz methanol maser survey are the
positions of central (proto)stars.
Class I methanol masers are generally located
offset from the central (proto)stars (e.g., Kurtz et al. 2004).
The peaks of such Class I methanol masers may
have offsets from the central objects, which would
affect the measured peak flux density and polarization properties.
A mispointing of $10\arcsec$
at 95~GHz ($15\arcsec$ at 44~GHz) can produce 1\% of artificial polarization
(See Section 2.2).
To solve this issue, we carefully adjusted target coordinates.
For the source with coordinates of central object, we made
9-point grid map. After the interpolation of the observed maser intensities
between the 9 points, we defined new coordinates if the offsets were
larger than $\sim 10\arcsec$. 
The sources included in the KVN VLBI fring survey (Kim et al. in prep.)
have refined positions for the 44~GHz methanol masers. The coordinates of
those sources were adopted from the fring survey.
In addition, we observed five points around the source with 
$30\arcsec$ spacing before each polarization measurement.
After similar interpolation of
intensities between the 5 points, we subsequently calculated the az/el offsets
of the 44~Ghz maser peak from the original coordinates.
Then, we carried out the polarization observations at the interpolated peak positions.
We note that L379 and NGC7538 has large offsets $> 24\arcsec$, and we determined
their coordinates based on the 9-point grid maps.
The coordinates and their references used for the polarization measurements
are described in Table~\ref{tb1}.

\subsection{Observations and Calibration}
We observed the 39 sources at 44 and 95~GHz simultaneously 
in full polarization spectral mode.
The observed lines are the Class~I methanol  $7_0 - 6_1 A+$ (44.06943 GHz) 
and $8_0 - 7_1 A+$ (95.169463~GHz) maser transitions.
The observations were conducted
using the KVN 21~m telescope at the Yonsei station in the single-dish mode
from August to December in 2013. 
The beam sizes are 65\arcsec\, and 30\arcsec\, at 44 and 95~GHz, respectively
\citep{Lee:2011aa, Kim:2011aa}.
A digital filter and a FX type digital spectrometer were used as a backend.
The spectrometer can produce auto power spectra of four input signals 
and cross power spectra of two input pairs simultaneously, 
which enable us to derive full Stokes parameters (I, Q, U, V) 
from the two polarization pairs
\citep[][Byun et al. in prep]{Oh:2011aa}.
Each spectrum of the spectrometer has 4096 channels.
The digital filter and spectrometer were configured to have
a bandwidth of 64~MHz for each spectrum, which results in a velocity
coverage of 430~\kms and a velocity resolution of 0.1~\kms at 44~GHz.

We used the Walsh Position Switching mode~\citep{Mangum:2000aa}, 
which repeats several pairs of 6 seconds on/off integration after firing cal.
For data reduction, we used the python polarization data pipeline module
available in KVN. 
The instrumental cross talk and phase offset were corrected using
planets (Jupiter, Venus, or Mars) and Crab, which were observed
at least once a day as calibrators.
The final spectra at 44 and 95~GHz were reduced to have the same velocity 
resolution of 0.2~\kms. The typical rms levels ($ 1 \sigma$) 
of the observed spectra are 0.5~Jy and 1.2~Jy at 44 and 95~GHz, respectively.

The polarized intensity ($PI$) given here is $(Q^2+U^2)^{1/2}$, and the
given error is the standard deviation of the measurement sets.
Detection criterion is that $PI > 3 \sigma $. 
The measured position angle was derived by 
$\chi = \frac{1}{2} \arctan(\frac{U}{Q})+152^\circ$. 
Here the latter term is the absolute position angle of Crab,
which is known to be nearly constant near the brightest region 
from millimeter to X-ray wavelengths \citep{Aumont:2010aa}.

Details on the single-dish polarization observation and calibration 
using KVN will be presented in a separate paper (Byun et al. in prep.).
The procedures of data reduction in the KVN polarization system
are also briefly explained in \citet{Kang:2015ac}. We also briefly describe 
the process here.
The KVN digital spectrometer backend for the polarization observations
produces two single-polarization spectra and one complex cross-polarization
spectrum, which are associated with the four Stokes parameters, I, Q, U, and V,
as \citet{Sault:1996aa} mentioned.
To estimate the leakage, so called combined D-term,
from the total intensity to the cross-polarization
spectrum, we used unpolarized planets, for which Q = U = V = 0.
The D-term measured by Jupiter were about 4\% and 8\% at 44 and 95~GHz,
respectively.
To evaluate the errors of the D-term, we observed planets multiple times
in a day, and calibrated the unpolarized planets either using another planet
or using the same planet observed with some time interval in a day.
For the Jupiter-Jupiter, Venus-Venus, or Jupiter-Venus pairs,
the measured polarization fractions were less than 0.4\%
and 2\% at 44 and 95~GHz, respectively.
These values would be the upper limits for the errors of the D-term
because they include the thermal noise of measurements.
Thus, the sources with polarization fractions above those values can be
treated real not artificial.

The polarization angles of targets were corrected based on the polarization
angle of Crab observed on the same day. We have found that the polarization
angles of Crab stayed constant within $4^\circ$ at 44 and 95~GHz when they
were measured on the same day.

The leakage variation due to the offset of the source from the beam
center was tested in the KVN system using unpolarized sources, 3C 84, Jupiter,
Venus, and Mars at 43~GHz at offsets from 0\arcsec\, to 30\arcsec\, with
5\arcsec\, or 10\arcsec\, intervals.
The results showed that the fractions of artificial polarization due to the pointing
offsets are 1\% at 15\arcsec\, and 2\% at 30\arcsec, respectively.
The leakage at 86~GHz was also tested with Jupiter and Mars
at offsets from 0\arcsec\, to 15\arcsec\, with 5\arcsec\, interval,
and 1\% of artificial polarized emission at 10\arcsec\, offset was measured.

The distortion of antenna beam pattern with elevation can cause variation
of leakage and it adds polarization calibration errors, too.
The gain loss due to the distortion of the beam pattern is represented as a
gain curve of antenna. The KVN Yonsei telescope has flat gain curves and
symmetric beam patterns over the elevation ranage from $30^\circ$ to $60^\circ$,
where our polarization observations were conducted.
The antenna gain changes only 1\% and 3\% at 43 GHz and 86 GHz, respectively,
over this elevation range.
The beam widths and the beam squints are almost constant over the observed
elevation range. The antenna beam widths in azimuth and elevation directions
are less than 2\arcsec\, different from each other, and the beam squints are
less than 2\arcsec .5 both in azimuth and elevation at both 43 and 86~GHz.
Therefore, we expect that the measured D-term is not significantly affected
by the antenna beam pattern or beam squint.
We note that the detailed system performance of the KVN Yonsei telescope
are reported in the status report of KVN
$(http://radio.kasi.re.kr/kvn/status\_ report\_2014/home.html)$.

\section{Results}

We detected fractional linear polarization toward 23 (59\%)
of the 39 Class I methanol maser sources at 44 and/or 95~GHz. 
This detection rate is slightly smaller than that (71\%) of 
\citet{Wiesemeyer:2004aa}, who found polarization in 10 out of 14 Class I
maser sources at 85, 95, and/or 133~GHz. 
Table~\ref{tb2} summarizes the observational results.
At 44 and 95~GHz, 21 (54\%) and 17 (44\%) sources show linear
polarization, respectively. We emphasize that this is the first
detection of linear polarization of the 44~GHz methanol masers.
Fifteen (38\%) sources were detected at both frequencies. The rms
weighted means of the fractional linear polarization detected sources are
$2.7\pm0.3$\% and $4.8\pm0.1$\% at 44 and 95~GHz, respectively. 
Eight sources are
detected at only one transition. Among them, NGC2264, which is
detected at 95~GHz, has $2\sigma$ possible detection at 44~GHz with
about 1\% of fractional polarization. The polarization upper limits of
the rest 7 sources are above the measured mean values, implying that
non-detection is due to the sensitivity.
It is worth noting that the 95~GHz polarization detections of OMC2 and G82.58+0.20
could be affected by the artifacts of the system polarization, because
their polarization fractions are same or only slightly higher than the upper
limit (2\%) of the artificial polarzation of the system at 95~GHz.

\floattable
\begin{deluxetable}{lr}
\tablecaption{Linear Polarization Detection Rates\label{tb2}}
\tablehead{\colhead{Group} & \colhead{Number}}
\startdata
Total Detected & 23  (59\%) \\
44 &           21  (54\%) \\
95 &           17  (44\%) \\
44 and 95 &    15  (38\%) \\
44 only &      6   \\
95 only &      2   \\
Total Observed & 39\\
\enddata
\end{deluxetable}

\begin{figure}
\plotone{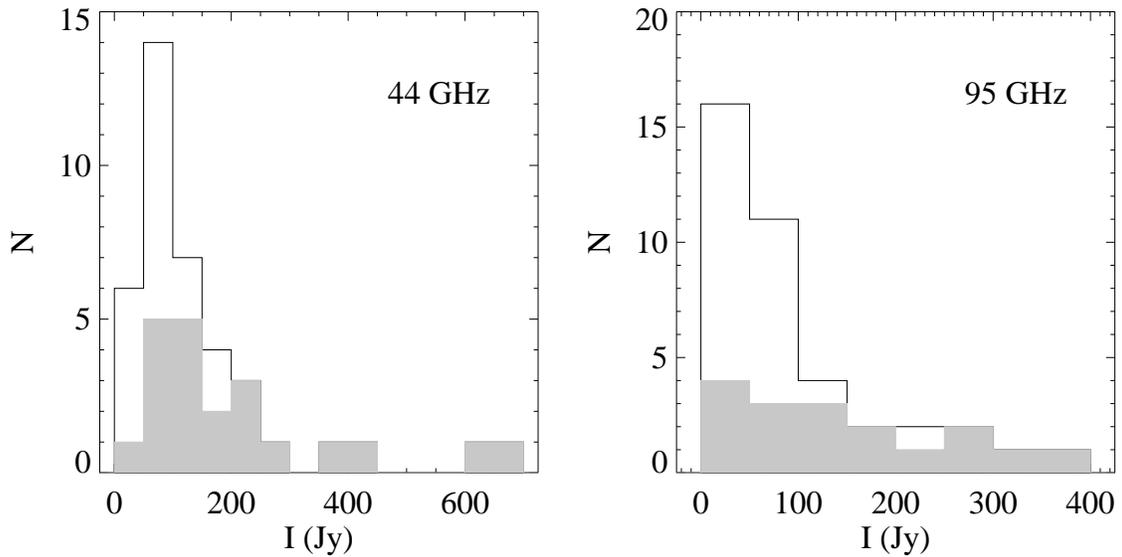}
\caption{Number density distributions of the total intensity (Stokes I) of
the observed (solid lines) and the polarization-detected sources (grey filled
area) at 44 (left) and 95~GHz (right).~\label{fg1}}
\end{figure}

\begin{figure}
\plotone{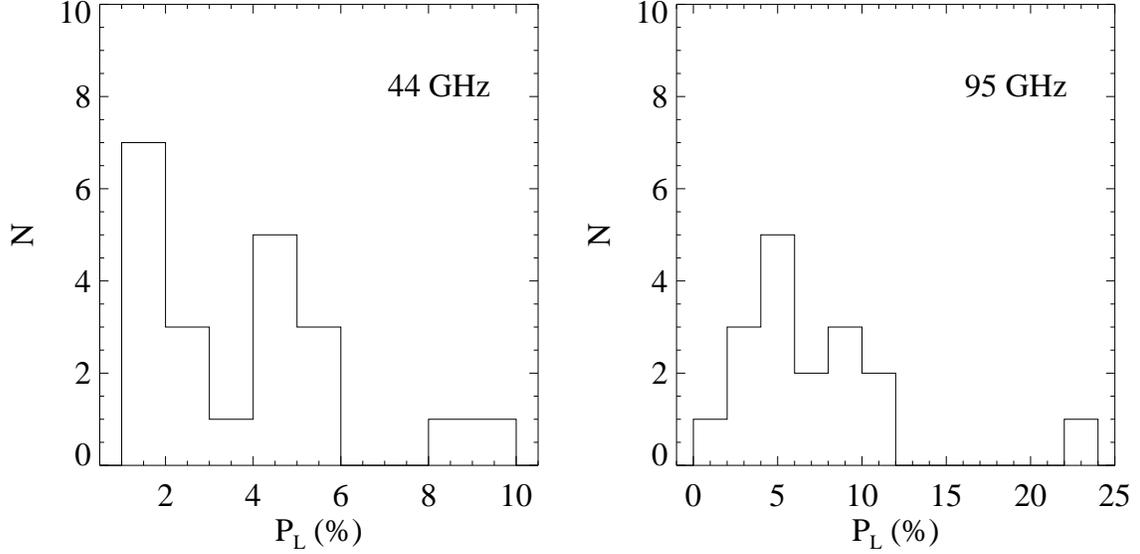}
\caption{Number density distributions of the polarization fractions
at 44 (left) and 95~GHz (right).~\label{fg2}}
\end{figure}

\begin{figure}
\plotone{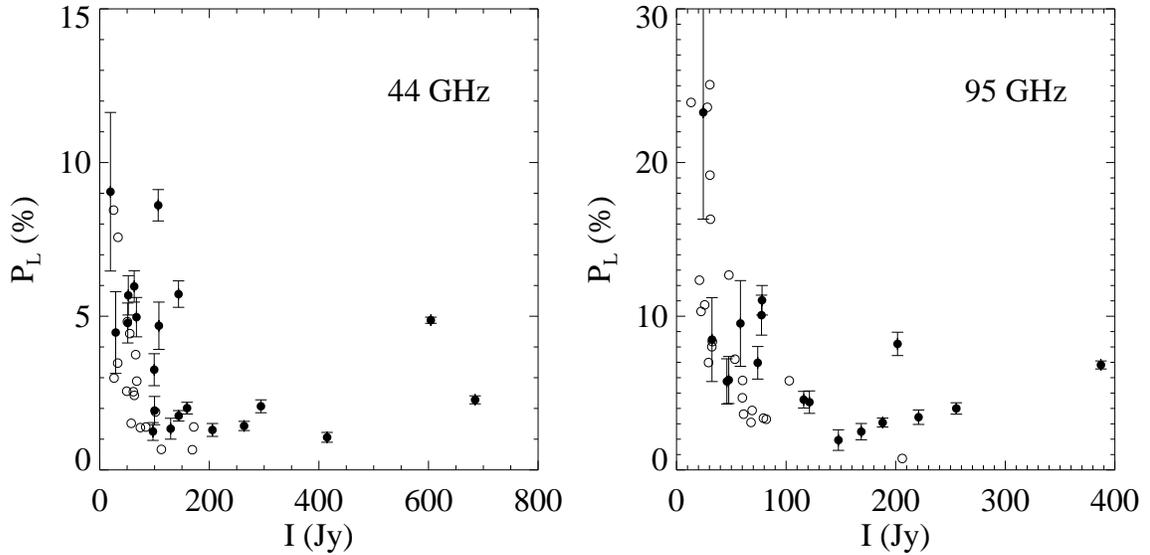}
\caption{Polarization fraction $P_L$ vs. Stokes I
at 44 (left) and 95~GHz (right). The filled circles are those
detected above $3\sigma$ errors, while open circles show
the upper limits of polarization non-detected sources.
Errors of $1\sigma$ are presented.~\label{fg3}}
\end{figure}

Figure~\ref{fg1} shows the number density distribution as a function of 
the total flux for the observed and the polarization detected sources.
The polarization detection rate tends to increase with the total flux 
at both transitions.
Figure~\ref{fg2} presents the distributions of fractional linear
polarization, $P_L$, 
at 44 and 95~GHz. All sources have $P_L < 11$\% except
W33Met (24.6\%), which has a large error of 6.9\% $(1\sigma)$.
The ranges of polarization fractions are
1.1\% -- 9.5\% and 2.0\% -- 24.6\% at 44 and 95~GHz, respectively. 
The ranges of upper limits for non-detections are
0.6\% -- 8.5\% at 44~GHz and 2\% -- 25\% at 95~GHz.
Figure~\ref{fg3} displays 
the polarization fraction distributions as a function of total flux. 
In this figure, the error of $P_L$ increases as $I$ decreases,
because $P_L \propto I^{-1}$ while the observational noise
is relatively constant for all sources
($\sim 0.5$~Jy at 44 and $\sim 1.2$~Jy at 95~GHz).
The upper limits of polarization non-detected sources are presented
with open circles.

Figure~\ref{fig1} presents the spectral profiles of the total flux $(I)$,
the polarized intensity $(PI)$, the polarization degree ($P_L$),
and the polarization position angle 
measured counterclockwise from north ($\chi$) of the sources detected
at both transitions.
The profiles of the total flux and the polarized intensity tend to peak
at similar velocities with some exceptions like G40.25-0.19
(Fig.~\ref{fig1}$l$ at 44~GHz). Some sources, e.g., G18.34+1.78SW
(Fig.~\ref{fig1}$f$), have multiple velocity
components with different polarization properties.
Figure~\ref{fig2} and  Figure~\ref{fig3} show
the same parameters for the sources detected only at 44 or 95~GHz, respectively.

The measured polarization properties are summarized in
Table~\ref{tb3}.
The presented values of $F_\nu$, $P_L$, $\chi$, and \vlsr 
are all determined in the channel with the peak polarized intensity.
The amplitude or the phase of a vector does not have a Gaussian probability
distribution unless the signal-to-noise ratio is very large.
This can lead to systematic errors in the measured polarizations.
According to \citet{Wardle:1974aa},
the best estimates of the true polarization
can be found using
$R \sim {R_M}^\prime [ 1 - (\sigma^\prime / {R_M}^\prime)^2]^{1/2}$
for $R/\sigma^\prime > 0.5$,
where ${R_M}^\prime$ is the observed polarization and $\sigma^\prime$
is the observed random noise.
Random noise does not produce a bias in $\chi$
but increases the error of $\chi$ 
by $\sigma_\chi \simeq \sigma^\prime / {R_M}^\prime$ radians.
We corrected for these effects in the presented $P_L$ and $\chi$.
The amount of correction in $P_L$ is insignificant in our cases, but $\sigma_\chi$
normally exceeds the measurement error. Errors presented are $1 \sigma$.

\begin{figure}
\plotone{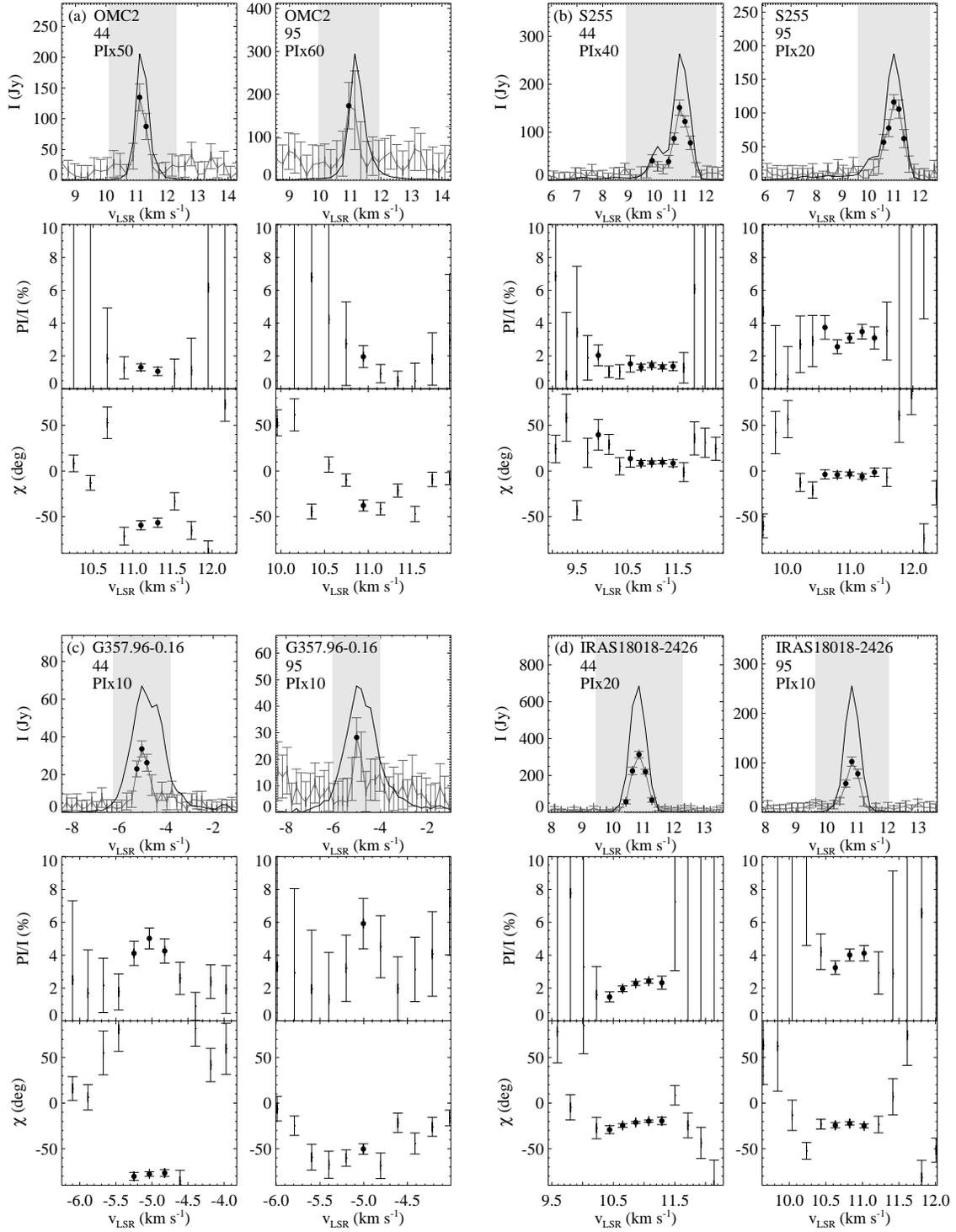}
\caption{Polarization-detected masers at both transitions.
The source name and line frequency are indicated in each panel.
For each source,
the polarized intensity $(PI)$ in grey line with errors (1$\sigma$) and
the total flux intensity $(I)$ in black solid line (top), and
the polarization fraction $P_L$ (middle) and angle $\chi$ (bottom)
are presented.
The multiplication factor of $PI$ is presented under the line frequency
for each source.
The points with $PI  > 3 \sigma$ are indicated with filled circles.
The grey-shaded area in the top panel indicates the velocity range for
the middle and bottom panels.\label{fig1}}
\end{figure}
\begin{figure}
\figurenum{4}
\plotone{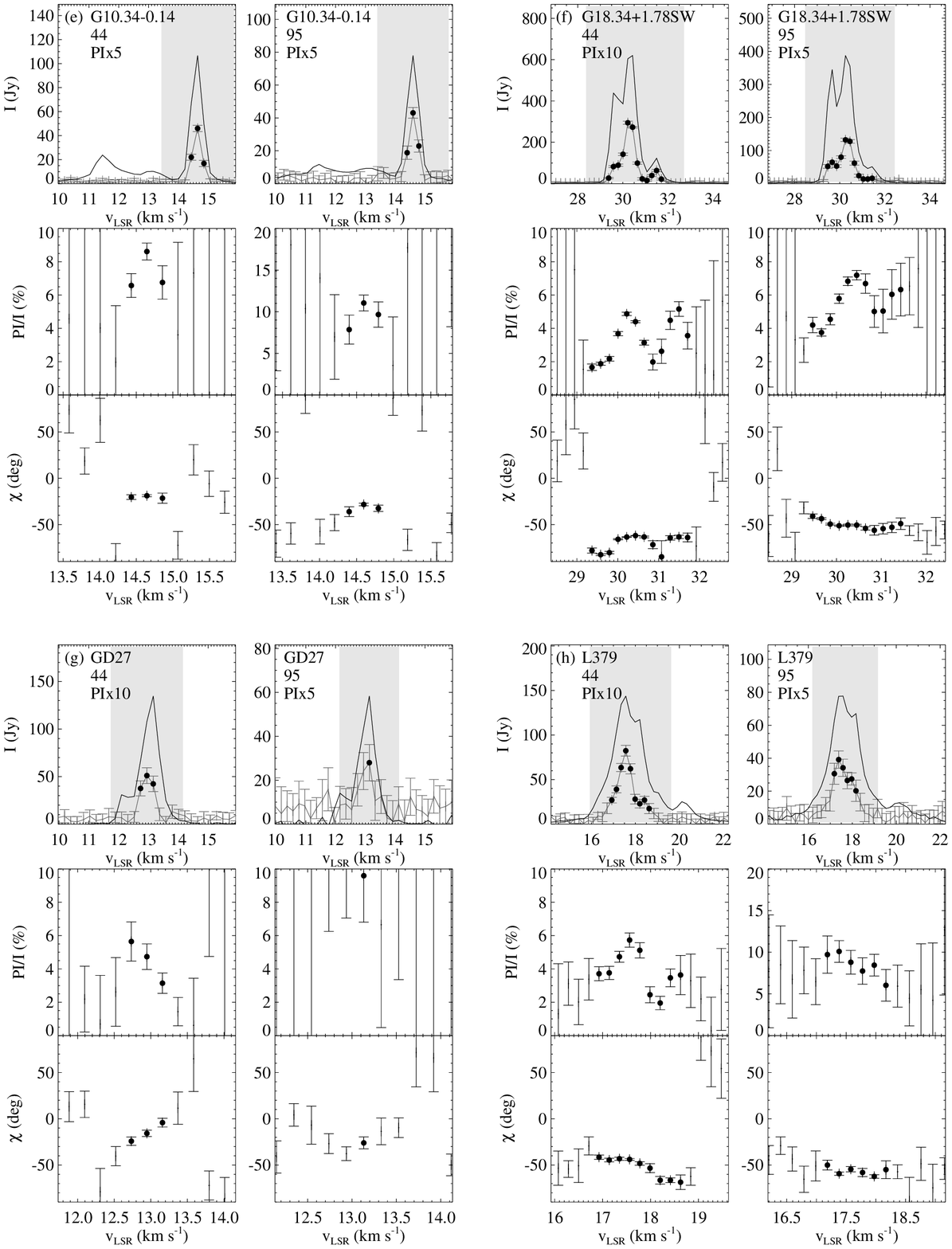}
\caption{(continued)}
\end{figure}
\begin{figure}
\figurenum{4}
\plotone{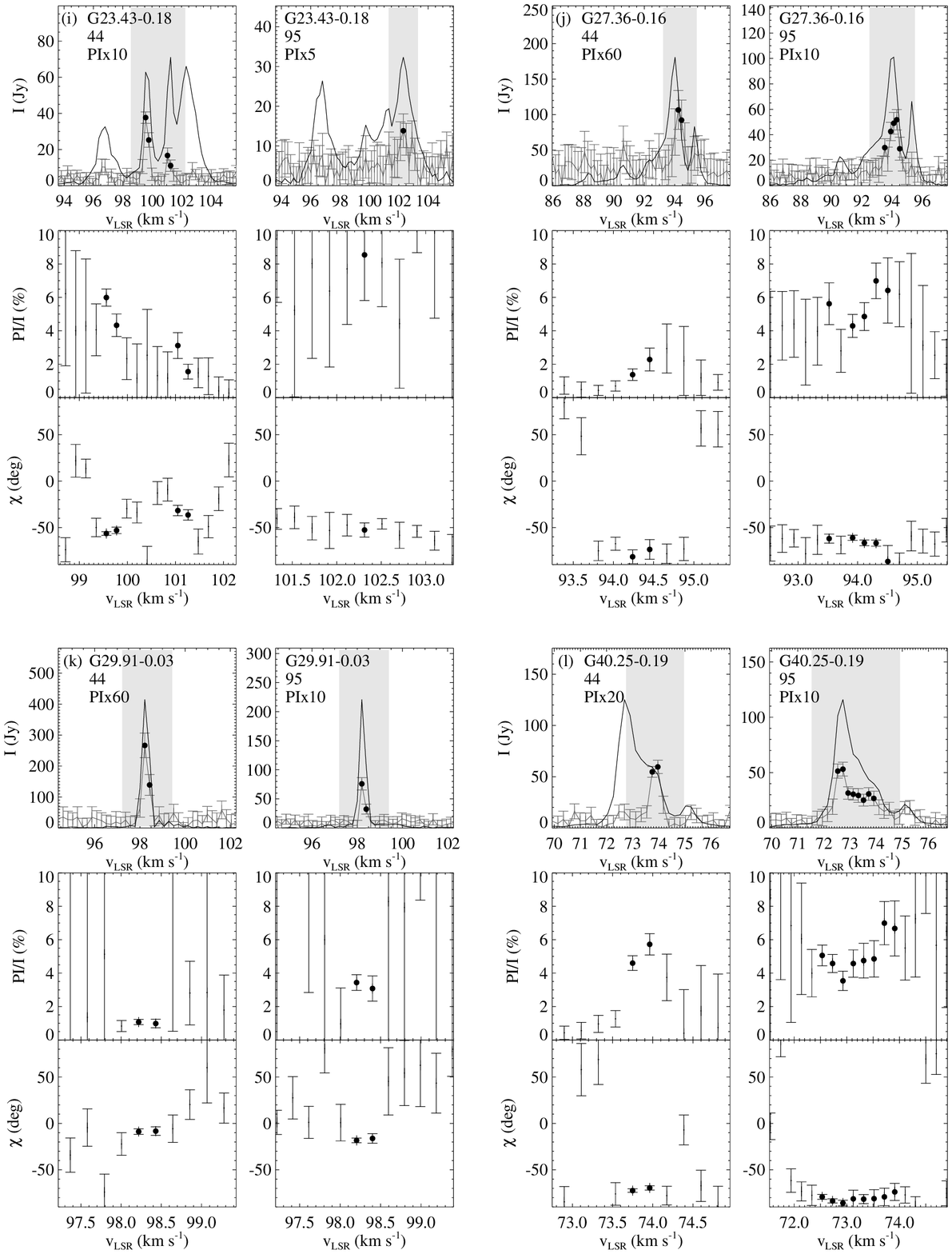}
\caption{(continued)}
\end{figure}
\begin{figure}
\figurenum{4}
\plotone{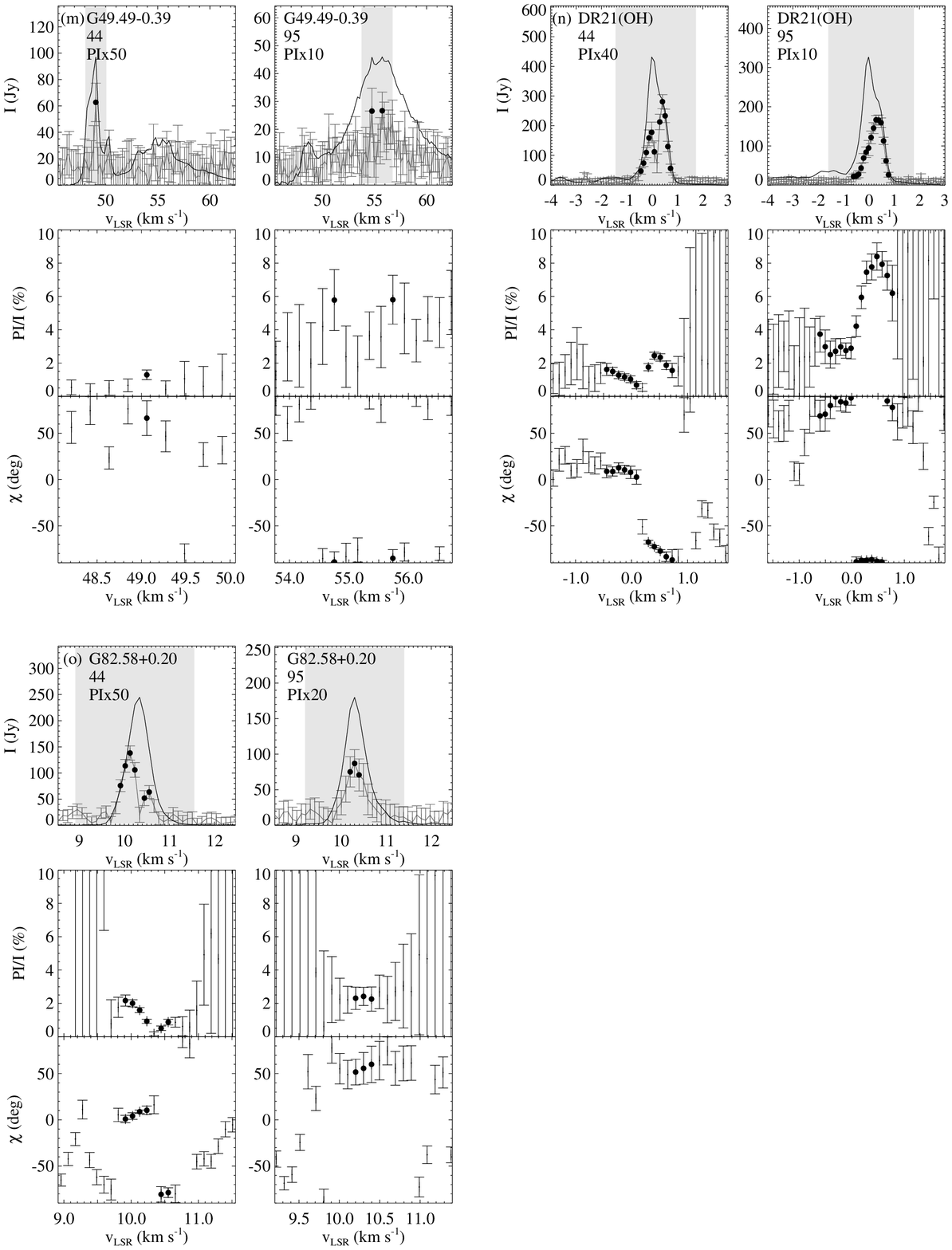}
\caption{(continued)}
\end{figure}

\begin{figure}
\plotone{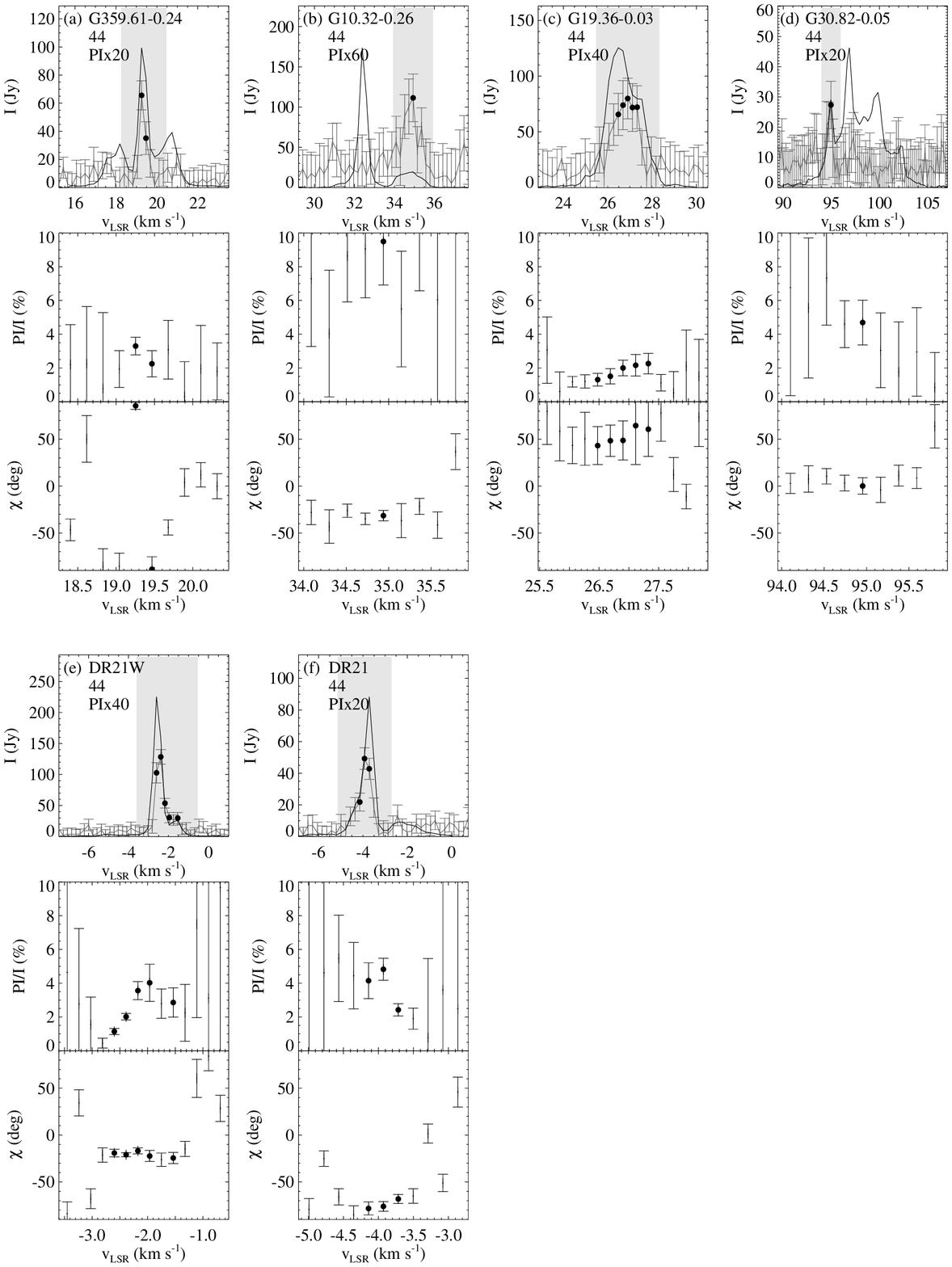}
\caption{Polarization-detected sources only at 44~GHz.\label{fig2}}
\end{figure}

\begin{figure}
\epsscale{0.5}
\plotone{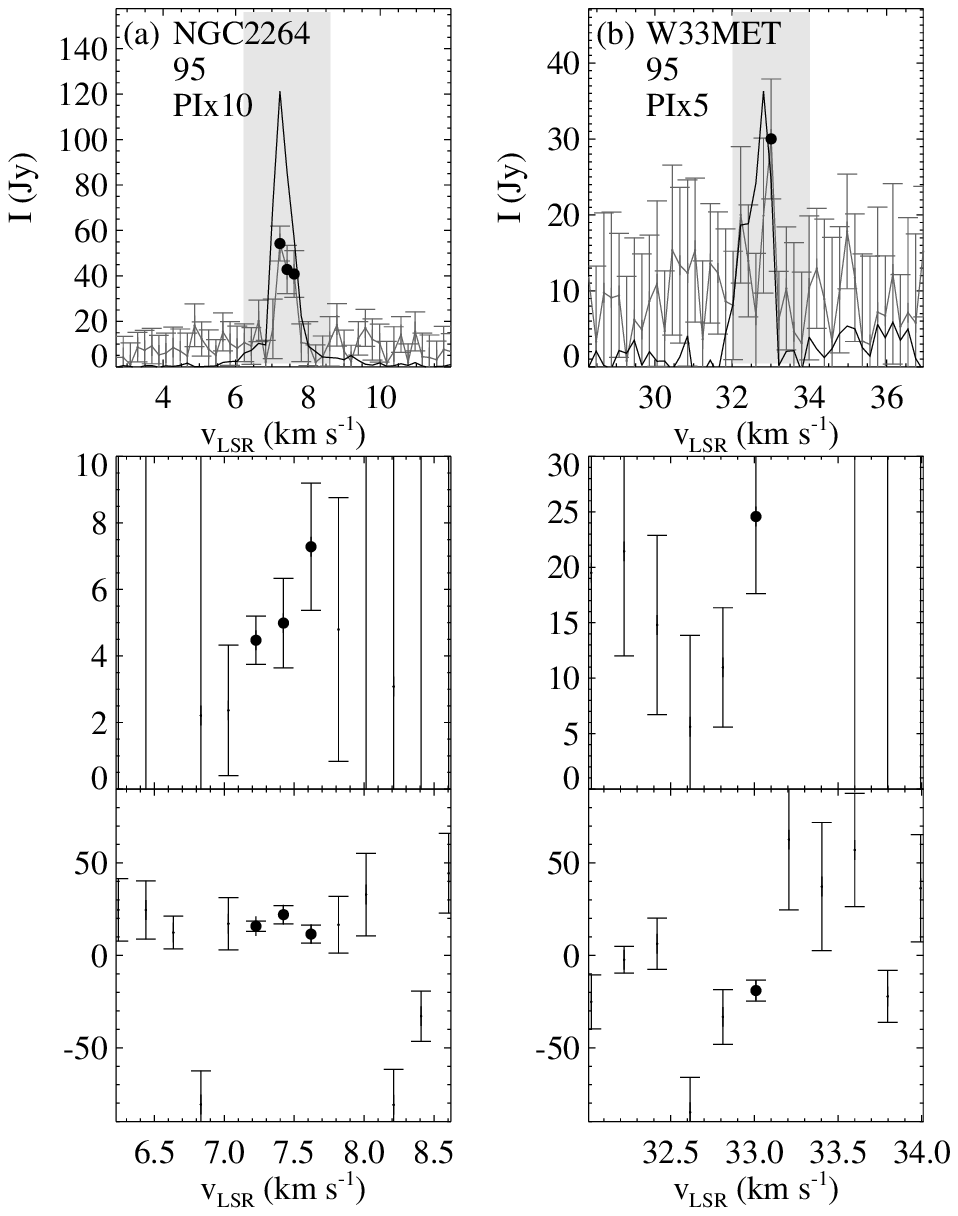}
\caption{Polarization-detected sources only at 95~GHz.\label{fig3}}
\end{figure}

\floattable
\begin{deluxetable}{llcrrrr}
\tablecaption{Linear Polarization Properties~\label{tb3}}
\tablehead{\colhead{Source} & \colhead{Transition} & \colhead{$F\nu$} &
\colhead{$P_L$}& \colhead{$\chi$} & \colhead{$v_{\rm LSR}$} \\
\colhead{Name} & \colhead{( GHz )} & \colhead{( Jy )} & \colhead{( \% )} &
\colhead{( $^\circ$ )} &\colhead{( {\rm km s$^{-1}$} )} }
\startdata
 OMC2 &44 &$205.6\pm2.8$ &$1.3\pm0.4$ &$120\pm9$ &$ +11.1$\\
  &95 &$147.8\pm5.1$ &$2.0\tablenotemark{a}\pm0.7$ &$142\pm8$ &$ +10.9$ \\
 S255N &44 &$263.4\pm0.8$ &$1.4\pm0.3$ &$9\pm5$ &$ +11.0$ \\
  &95 &$188.1\pm0.7$ &$3.1\pm0.3$ &$176\pm4$ &$ +11.0$ \\
 NGC2264 &95 &$121.3\pm4.0$ &$4.5\pm0.9$ &$15\pm10$ &$  +7.2$ \\
 G357.96-0.16 &44 &$67.0\pm0.4$ &$5.0\pm0.6$ &$102\pm8$ &$  -5.0$ \\
  &95 &$47.7\pm0.8$ &$5.9\pm1.5$ &$129\pm10$ &$  -5.0$ \\
 G359.61-0.24 &44 &$99.4\pm1.4$ &$3.3\pm0.5$ &$85\pm9$ &$ +19.3$ \\
 IRAS18018-2426 &44 &$685.0\pm5.5$ &$2.3\pm0.6$ &$158\pm3$ &$ +10.9$ \\
  &95 &$255.3\pm1.2$ &$4.0\pm0.6$ &$157\pm4$ &$ +10.8$ \\
 G10.34-0.14 &44 &$106.7\pm0.7$ &$8.6\pm0.5$ &$161\pm3$ &$ +14.6$ \\
   &95 &$78.0\pm1.8$ &$11.0\pm1.0$ &$151\pm3$ &$ +14.6$ \\
 G10.32-0.26 &44 &$19.6\pm0.4$ &$9.5\pm2.6$ &$148\pm18$ &$ +34.9$ \\
 W33MET &95 &$24.4\pm1.5$ &$24.6\pm6.9$ &$160\pm19$ &$ +33.0$ \\
 G18.34+1.78SW &44 &$604.2\pm5.7$ &$4.9\pm0.5$ &$116\pm1$ &$ +30.2$ \\
   &95 &$387.3\pm1.9$ &$6.8\pm0.5$ &$129\pm2$ &$ +30.3$ \\
 GGD27 &44 &$107.9\pm1.0$ &$4.7\pm0.8$ &$164\pm8$ &$ +12.9$ \\
  &95 &$58.3\pm3.6$ &$9.6\pm2.8$ &$153\pm7$ &$ +13.1$ \\
 G19.36-0.03 &44 &$99.9\pm0.7$ &$2.0\pm0.5$ &$48\pm21$ &$ +26.9$ \\
 L379 &44 &$143.8\pm1.8$ &$5.7\pm0.5$ &$136\pm4$ &$ +17.6$ \\
  &95 &$77.6\pm3.1$ &$10.1\pm1.3$ &$120\pm5$ &$ +17.4$ \\
 G23.43-0.18 &44 &$62.9\pm0.8$ &$6.0\pm0.5$ &$123\pm6$ &$ +99.6$ \\
 &95 &$32.3\pm1.1$ &$8.5\pm2.7$ &$127\pm8$ &$+102.3$ \\
 G27.36-0.16 &44 &$129.4\pm2.5$ &$1.4\pm0.4$ &$98\pm13$ &$ +94.2$ \\
  &95 &$74.1\pm0.8$ &$7.0\pm1.1$ &$113\pm5$ &$ +94.3$ \\
 G29.91-0.03 &44 &$414.9\pm2.7$ &$1.1\pm0.5$ &$171\pm6$ &$ +98.2$ \\
 &95 &$220.9\pm4.0$ &$3.4\pm0.5$ &$161\pm4$ &$ +98.2$ \\
 G30.82-0.05 &44 &$29.1\pm0.6$ &$4.7\pm1.3$ &$0\pm18$ &$ +95.0$ \\
 G40.25-0.19 &44 &$52.0\pm0.6$ &$5.7\pm0.6$ &$110\pm8$ &$ +74.0$ \\
 &95 &$116.0\pm3.5$ &$4.6\pm0.5$ &$96\pm4$ &$ +72.7$ \\
 G49.49-0.39 &44 &$97.1\pm2.0$ &$1.3\pm0.3$ &$66\pm19$ &$ +49.1$ \\
 &95 &$46.0\pm0.7$ &$5.8\pm1.5$ &$94\pm10$ &$ +55.7$ \\
 DR21W &44 &$159.3\pm1.7$ &$2.0\pm0.2$ &$159\pm4$ &$  -2.4$ \\
 DR21(OH) &44 &$294.1\pm2.9$ &$2.1\pm0.2$ &$109\pm3$ &$  +0.4$ \\
  &95 &$201.7\pm5.7$ &$8.2\pm0.8$ &$92\pm2$ &$  +0.4$ \\
 DR21 &44 &$51.0\pm1.3$ &$4.8\pm0.7$ &$103\pm8$ &$  -3.9$ \\
 G82.58+0.20 &44 &$144.0\pm1.3$ &$1.8\pm0.2$ &$6\pm5$ &$ +10.1$\\
 &95 &$168.6\pm3.2$ &$2.5\tablenotemark{a}\pm0.5$ &$55\pm16$ &$ +10.3$ \\
\enddata
\tablenotetext{a}{We note that the 95~GHz polarization fractions of OMC2
(2.0\%) and G82.58+0.20 (2.5\%) are same or close to the upper limit (2\%) of
the artificial polarization of the system at 95~GHz. Thus, we cannot discard
the possibility that their detections were affected by the artifacts of the
system.}
\tablecomments{
The presented values of $F_\nu$, $P_L$, $\chi$, and \vlsr 
are all determined in the channel with the peak polarized intensity.}
\end{deluxetable}

\begin{figure}
\plotone{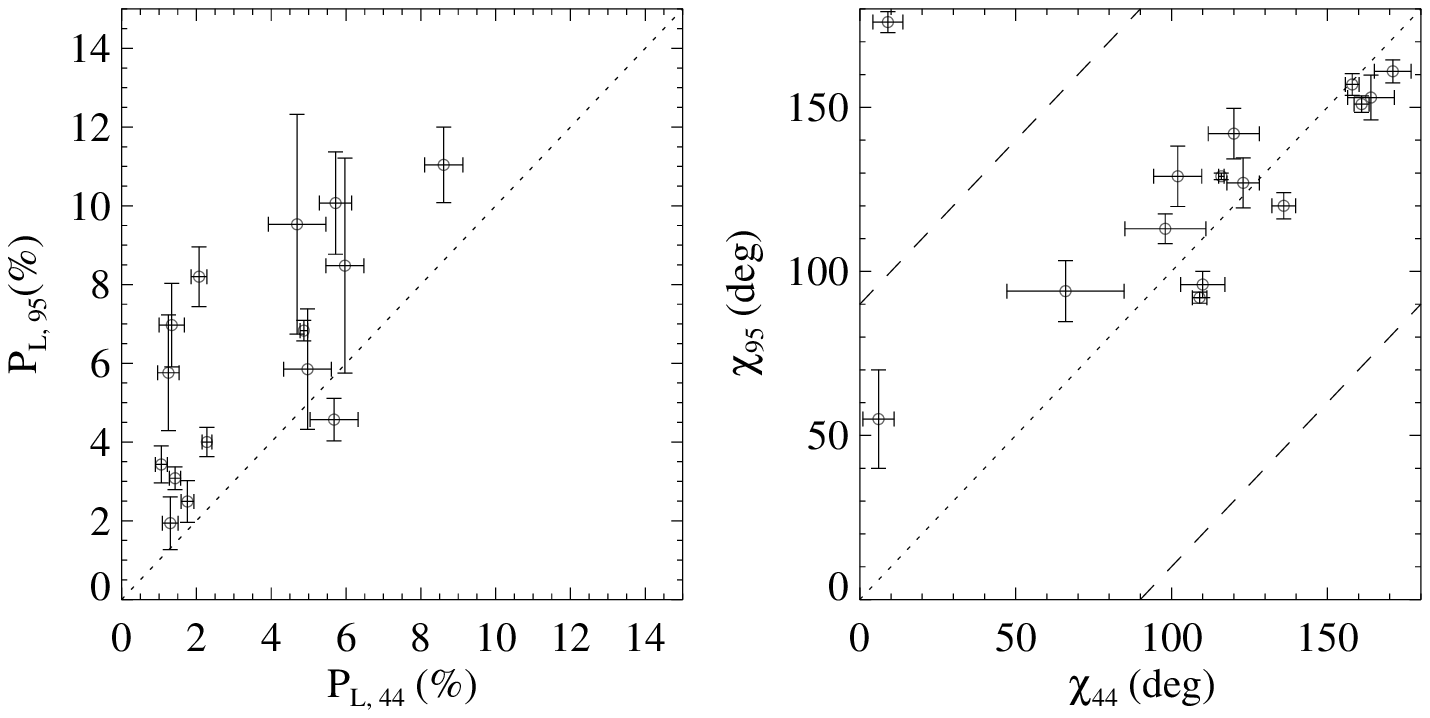}
\caption{44 GHz vs. 95 GHz masers in polarization fraction, $P_L$ (left)
and angle, $\chi$
(right) for the sources detected in both transitions at the same velocities.
Errors of $1\sigma$ are presented.
The dotted lines
show the cases when the polarization degree and angles of the two transitions
are the same.
The dashed lines in the right panel
show the cases when the angels are $90^\circ$ different.
~\label{fg4}}
\end{figure}

The polarization properties of the 44 and the 95~GHz methanol masers
are observed to be well correlated.
In Figure~\ref{fg4}, the polarization degrees of the two transitions
appear to have a positive linear correlation, although the correlation is
not very tight.
Their linear Pearson correlation coefficient is $r=0.71$,
where $r=0$ and $1$ indicate no correlation and a perfect positive
linear correlation, respectively.
The polarization fractions of the 95~GHz masers tend to be greater than those
of the 44~GHz masers. 
This is consistent with the finding of \citet{Wiesemeyer:2004aa},
where the polarization fraction at higher frequency transition
is higher than that at lower frequency whenever
a source is detected at multiple maser transitions.
Figure~\ref{fg4} compares the position angles of the 44 and
the 95~GHz maser transitions as well. Their correlation coefficient is $r = 0.92$,
indicating that their position angels are generally well correlated.

\section{Discussion}

\subsection{The Measured Polarization Properties}

About 40\% of the observed sources did not show any linear polarization
in either of the two transitions.
Because of the
observational limitations such as sensitivity and a large single dish beam
depolarizing multiple maser features with different orientations,
we can not exclude the presence of linear polarization even for
the non-detected sources.

The measured linear polarization fractions in this study are mostly less than
10\%. The fractions could be
lower limits, because there are several effects reducing the original
degree of linear polarization.
Linear polarization can be reduced by field line curvature
along the propagation direction by destroying the phase coherence of polarization \citep{Elitzur:2002aa},
or by internal Faraday rotation due to the presence of electrons in the magnetized medium
along the maser amplification path \citep{Fish:2006aa}. However,
the effects of internal Faraday rotation is expected to be small~\citep{Surcis:2011aa},
so our results are not likely affected by the Faraday rotation.
The fraction can also be reduced by
depolarization because of combining multiple maser features with different polarization
orientations within a large single dish beam.

None of the 5 sources detected at 84 GHz or 95 GHz by
\citet{Wiesemeyer:2004aa} were detected in our 95 GHz observations, although
some targets have high fractional polarizations up to $\sim 9$\%.
This is mainly because of the sensitivity issue.
Four (S231, W51Met2, W75S(3), and
NGC7538) were weak sources with total fluxes below 20 Jy, and could not be
detected in our 95 GHz observations with a typical rms of 1.2 Jy and the
$3 \sigma$ detection criterion.
The other (DR21W) was bright (210 Jy) but was rejected from the detection list
because it
had a polarization fraction slightly lower than the upper limit (2\%) of the
artificial system polarization at 95 GHz.

We have not detected any source with linear polarization above 30\%
that \citet{Wiesemeyer:2004aa} has reported at 132~GHz Class I and
157~GHz Class II maser transitions. The highest estimated linear polarization
degrees were 40\% and 37\% at 132 and 157~GHz, respectively.
Linear polarization above 33\% is expected to be rare~\citep{Elitzur:2002aa}. 
It is possible
only when the angle between the magnetic field and the ray path is
$35^\circ < \theta < 45^\circ$ for isotropic pumping.
\citet{Nedoluha:1990aa} indicated that, for an angular momentum $J = 1-0$
transition,
the fractional polarization as high as 100\% is observable
when a Zeeman frequency is large, however,
about 30\% would be the highest for an angular
momentum $J = 2-1$ and higher transitions, unless significant anisotropic
pumping is present.
\citet{Wiesemeyer:2004aa} found that a large fractional linear polarization
$(P_L >33\%)$ is not rare ( 2 out of 10 for Class I and 1 out of 3 for Class
II), giving an impression that
anisotropic pumping or loss may be commonly achievable due to an unequal
population of the magnetic substates of the maser levels~\citep{Nedoluha:1990aa}.
In the present study, none of 23 polarization detected sources show such
high linear polarization. 
We also observed 2 sources (M8E and L379)
that showed highest fractional polarization $> 30$\% at 132~GHz
in \citet{Wiesemeyer:2004aa}.
The two have fractional polarizations of 4\% -- 10\% at 95~GHz.
Since transitions at higher frequencies tend to have higher fraction of
polarization, we cannot directly compare our results with those of
\citet{Wiesemeyer:2004aa}.
However, our observational results, none with such high fractional polarization
out of 23 polarization detected sources, still
suggest that the anisotropic pumping
or loss mechanism may not be common for 44 and 95~GHz maser transitions.

\subsection{Comparison of Polarization Properties at 44 and 95~GHz}

We have found that the polarization fractions and angles of the 44 and 95~GHz
masers are generally well correlated as shown in Figure~\ref{fg4}.
Their fractions of linear polarization tend to have a linear correlation
and their polarization angles are aligned well.
It has been suggested that the two transitions trace similar regions
\citep[e.g.,][]{Valtts:2000aa, Kang:2015ab}.
The similar polarization properties of the two lines also
confirm that the masers at these two transition lines are indeed experiencing
magnetic fields of similar regions.

Our observations show that the degree of linear polarization at 95~GHz
is greater than that at 44~GHz for all 15 sources detected at both
frequencies with one exception (G40.25$-$0.19).
The reduced fractional polarization at the lower frequency transition
compared to higher frequency transitions seems to be general trend
in maser polarization observations. For example,
\citet{McIntosh:1993aa} and \citet{Wiesemeyer:2004aa} reported the same
trend for silicon monoxide and methanol masers, respectively.
Several possibilities may be suggested to explain these observations.

The Faraday rotation is suggested as an explanation of depolarization
at longer wavelengths \citep{Elitzur:2002aa}.
The effects of Faraday rotation on the linear polarization
of the masers are well described by
\citet{Fish:2006aa} and \citet{Surcis:2011aa}.
According to their explanation, internal Faraday rotation
due to the free electrons in the path of maser amplification 
can decrease the linear polarization fraction of
the radiation, sometimes completely circularizing it if
the Faraday rotation is strong enough~\citep{Goldreich:1973aa}.
However, \citet{Surcis:2011aa} show that the depolarization
due to the internal Faraday rotation
for the 6.7~GHz methanol maser is negligible in case of C-shock,
and it is not likely significant in case of J-shock either.
The rotation measure changes due to the external Faraday rotation is
negligibly small to produce depolarization.
When $RM = 1100~{\rm rad~m^{-2}}$, the maximum value
measured toward pulsars in \citet{Han:2006aa}, is adopted,
the angle due to the Faraday rotation is only $3^\circ$ at 44~GHz,
which in general smaller or comparable to the angle measurement error.
Thus, the Faraday rotation can not explain the lower polarization fraction
at 44~GHz.

Another possibility is the lower $J$ transition data averaging over a larger
volume of medium due to the larger beam than the other transitions
\citep{McIntosh:1993aa}. Variation of physical conditions in a larger
volume may reduce the fractional polarization. However, the similarities
in total flux and polarization properties in our observations
suggest that both transitions are likely covering similar maser components.
In addition, according to the previous
VLA observations~\citep{Kogan:1998aa, Kurtz:2004aa} of methanol masers,
a case of
the 44~GHz masers being distributed over area wider than the KVN beam,
$30^{''}$ at 95~GHz, is rare. Thus, different volume does not seem to
be the main cause of the difference of fractional polarization.

The fractional linear polarization of the 44 and 95~GHz transitions
could be intrinsically different.  \citet{Perez-Sanchez:2013aa}
calculated the fractional linear polarization as a function of the
emerging brightness temperature for some rotational transitions of
SiO, H$_2$O and HCN masers.  In their calculation, the higher $J$
transitions have somewhat lower polarization percentages, which is
mainly due to the averaging over more magnetic substates.  However,
this assumes similar levels of stimulated emission rate, $R$. As also
shown in \citet{Nedoluha:1990aa}, the linear polarization fraction is
a function of the degree of maser saturation $R/\Gamma$, a ratio of
stimulated emission rate to decay rate, and the angle $\theta$ between
the line of sight (LOS) and the magnetic field orientation. One of
their main finding is that the polarization characteristics for high $J$
transitions are similar in quality and quantity, suggesting the
behaviors of the $7_0 - 6_1 A+$ and $8_0 - 7_1 A+$ transitions would
be similar as long as they have similar saturation level and $\theta$.
The angle $\theta$ is likely to be similar for both transitions.
Although not much is known about the decay rate of the
masers~\citep{Vlemmings:2010aa}, in our case, $R$ is also similar for both
maser transitions.  The stimulated emission rate $R$ is given by $R
\sim \frac{AkT_b \Delta \Omega}{4\pi h \nu}$, where A is the Einstein
coefficient of the involved transition, $k$ and $h$ are the Boltzmann
and Planck constant, $\nu$ is the maser frequency, $T_b$ is the
brightness temperature, and $\Delta \Omega$ is the relation between
the real angular size of the masing cloud and the observed angular
size \citep{Perez-Sanchez:2013aa}.  The Einstein coefficients are
$A_{\rm 44~GHz}\sim2.74\times 10^{-7}$~s$^{-1}$, and $A_{\rm
95~GHz}\sim2.87\times 10^{-6}$~s$^{-1}$ (Lankhaar et al., 2016, in prep.).
The brightness temperature $T_b$ can be estimated using $T_b = S_\nu
\Sigma^{-2} \zeta_\nu$, where $S_\nu$ is the observed flux density,
$\Sigma$ is the maser angular size, and $\zeta_\nu$ is a constant
factor that includes a proportionality factor obtained for a Gaussian
shape. This factor $\zeta_\nu \propto \nu^{-2}$, and is
$\sim3.16\times10^8$ at 44~GHz and $\sim6.77\times10^7$ at 95~GHz
respectively \citep{Perez-Sanchez:2013aa}. Thus, $T_b$ is roughly five
times larger at 44~GHz than at 95~GHz when assuming the same physical
size and flux. When additionally the same beaming angle is assumed,
the stimulated emission rate $R_{\rm 44~GHz}\approx0.96~R_{\rm
95~GHz}$. As the flux of the 95~GHz masers is found to be
$\sim50\%-80\%$ of that of the 44~GHz masers, $R_{\rm 44~GHz}$ for the
individual masers is unlikely to be more than $\sim2\times R_{\rm 95~GHz}$. 
Saturation effects are thus unlikely to contribute to the
observed differences unless the physical sizes of the masers and/or
the beaming angles of the two transitions are very different. Related
to this, the dependence of the linear polarization on the ratio
between the Zeeman splitting rate and the stimulated emission rate,
$g\Omega/R$, can still contribute to the observed differences between
the two maser transitions \citep{Nedoluha:1990aa}, provided $g\Omega$
is sufficiently different between the two transitions. Full
understanding on the intrinsic cause of higher polarization fraction
in higher frequency transition line seems to need supports from the
theoretical studies in the future.

\subsection{Association with Outflows or Galactic Scale Magnetic Fields}

Class I methanol masers are collisionally excited and thought to occur
in the areas where the methanol molecular abundance is enhanced by
the outflow shock heating of grain mantles~\citep{Menten:1991aa}.
If they arise in the regions compressed by outflow shock,
then the observed polarization angles could be associated with the
orientations of outflows.
We have searched for the references related to the outflows of
the 23 polarization detected sources 
using the SIMBAD website~\citep{Wenger:2000aa}.
Many sources have references indicating existence of outflows,
such as H$_2$ maps, extended green objects,
or outflow tracing molecular line observations, 
but directions of outflows are not obvious in many cases.
Among them,
we have found 7 sources,
i.e., OMC2, S255N, NGC2264, G49.49$-$0.39 (W51 e2), DR21W, DR21, and DR21(OH),
where the direction of maser-associated outflow is rather simple.
IRAS18018-2426, GGD27, L379, and G30.82-0.05 (W42 main) have been mapped 
in CO transition lines as well,
but the connection between the maser and outflow is not obvious 
either due to the outflow direction being in the line-of-sight 
or due to the complicated multiple outflows morphology
\citep{Zhang:2005aa, Fernandez-Lopez:2013aa, Kelly:1996aa, Sridharan:2014aa}.

We have compared the polarization angles at 44~GHz for 6 sources 
and at 95~GHz for NGC2264, which are not detected at 44~GHz, with
the orientations of outflows available in the literatures.
The position angles (PA) of their outflows, their references,
and the angles between the polarization angle and the outflow
are summarized in Table ~\ref{tb4}. 
Regarding the errors of $| \chi - PA_{\rm out}|$, we took the measurement
errors of the polarization angles, but the errors would be larger
considering that the errors of outflows
are normally larger. For example, the conventional error of outflow is
considered to be $15^\circ$ in \citet{Surcis:2013aa}.
There are some sources with their polarization
angles being almost perpendicular to the outflow direction, e.g., OMC2 and
DR21W. However, in general, the association between
outflows and polarization angles is not apparent.

\begin{figure}
\plotone{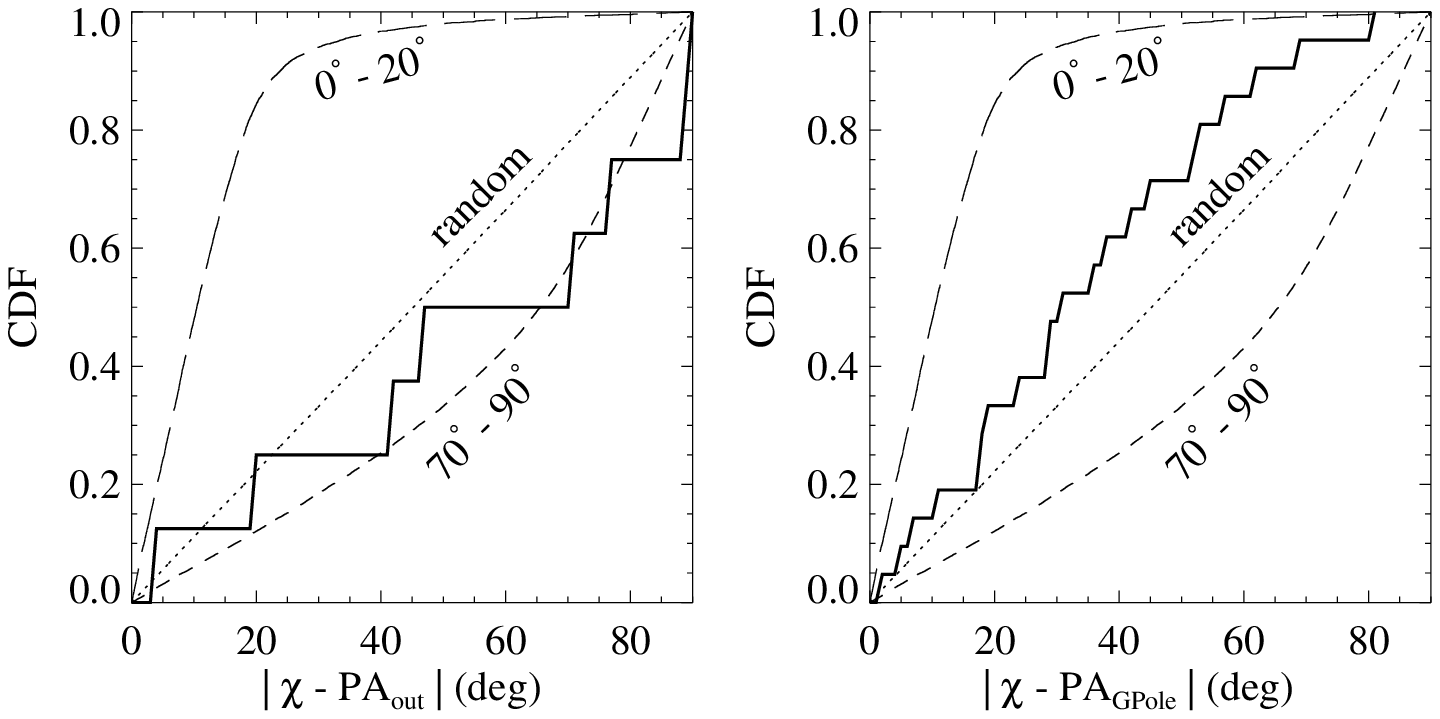}
\caption{Thick solid lines show the cumulative distribution functions
(CDF) of the angles between the polarization angles and the orientations
of outflows (left) and those of the Galactic pole (right).
The dotted line and dashed and long-dashed curves are the CDFs of projected
angles from Monte Carlo simulations where two vectors are
randomly distributed, perpendicular, and well-aligned to each other,
respectively.~\label{fig45}}
\end{figure}

Although the sample size is small, we tried the Kolmogorov-Smirnov (K-S)
statistics to see whether the angle differences are similar to
or different from the projected
angles of aligned, perpendicular, or randomly aligned samples,
by comparing the observed angle difference to the results from the
Monte Carlo simulations as discussed in \citet{Hull:2014aa}.
The simulation that randomly selects pairs of vectors in three dimensions
gives the same distribution when it is projected on the plane of the sky.
However, the projection effects increase the scatter of 2-d projected angles
in other cases.
For the aligned case, we select pairs of vectors that are aligned within
$20^\circ$ of one another among the random pairs in three dimensions
and calculate their angles projected onto the plane of the sky.
For the perpendicular case, we select pairs of vectors separated by $70^\circ
- 90^\circ$, and projected them. The projection effects
are more important in the perpendicular case then those in the aligned case.
The cumulative distribution functions (CDF) of the 2-d projected angles
from simulations are compared with the CDF of the
observed angle difference between the outflow direction and the 
polarization angle in the left panel of Figure~\ref{fig45}.

\floattable
\begin{deluxetable}{lrrrr}
\tablecaption{Outflow Orientations and Polarization Angles\label{tb4}}
\tablehead{
\colhead{Source Name} & \colhead{$\chi$} & \colhead{$PA_{\rm out}$} & 
\colhead{$| \chi - PA_{\rm out} |$}&
\colhead{Reference}}
\startdata
OMC2 & $120^\circ$ & $30^\circ$ & $90^\circ \pm 9^\circ$ & \citet{Williams:2003aa} \\
S255N & $9^\circ$ & $51^\circ$ & $42^\circ \pm 5^\circ$ & \citet{Wang:2011aa} \\
NGC2264 & $15^\circ$ & $175^\circ$ & $20^\circ \pm 10^\circ$ & \citet{Schreyer:2003aa} \\
G49.49$-$0.39 (W51 e2) & $66^\circ$ & $137^\circ$ & $71^\circ \pm 19^\circ$ & \citet{Shi:2010aa}\\
DR21W & $159^\circ$ & $70^\circ$ & $89^\circ \pm 4^\circ$ & \citet{Garden:1991aa}\\
DR21(OH) & $109^\circ/36^\circ$ & $113^\circ$ & $4^\circ \pm 3^\circ /77^\circ \pm 3^\circ $ & \citet{Zhang:2014aa}\\
DR21 & $103^\circ$ & $150^\circ$ & $47^\circ \pm 8^\circ$ & \citet{Garden:1991aa}\\
\enddata
\tablecomments{$\chi$ at 44~GHz is used for comparison except NGC2264,
which is polarization detected only at 95~GHz.
Position Angels of outflows (PA$_{\rm out}$) are estimated
counterclockwise from the north, same as the way that the polarization
angles of 44~GHz masers are measured. The reference for the outflow orientation
is noted in the Reference column.
The error of the PA difference in column 4 is adopted from the
44~GHz maser polarization angle measurement error.
DR21(OH) has two perpendicular polarization angles, resulting in two
perpendicular $| \chi - PA_{\rm out} |$ values.}
\end{deluxetable}

The probabilities of the observed $| \chi - PA_{\rm out} |$
and the simulated projected angles being different or similar derived
from the K-S tests are given in Table~\ref{tb5}.
The K-S tests rule out the scenario where the outflows and polarization angles
are tightly aligned $(P < 0.01)$. The probability of them being
perpendicular $(P  = 0.7)$ appears to be higher than that of them being
random $(P  = 0.4)$. Magnetic fields could be either parallel or perpendicular
to the polarization angle depending on the angle between
the magnetic field and the ray path, $\theta$. If we simply assume
that methanol masers are strongly saturated $( R/ \Gamma > 10^2)$
because they are strong, then the polarization angle is perpendicular
to the magnetic field regardless of $\theta$~\citep{Nedoluha:1990aa}.
If we perform the K-S tests for 
$| B - PA_{\rm out} |$ taking into account above assumption,
the K-S tests returns $P = 0.5$ for random, and $P < 0.1$ for aligned and
perpendicular cases of simulations.
Considering the ambiguity of magnetic field direction derived from
the polarization angle, limited number of
samples, and the difficulty of distinguishing the projected angle
distribution from the random and $70^\circ -90^\circ$ simulations,
we cannot conclude whether magnetic fields/polarization angles
are randomly aligned or mis-aligned with outflows, however,
we can rule out the scenario of them being tightly aligned at least.

\floattable
\begin{deluxetable}{lrr}
\tablecaption{Probability of the K-S Test~\label{tb5}}
\tablehead{ \colhead{} & \colhead{$| \chi - PA_{\rm out} |$} & 
\colhead{$| \chi - PA_{\rm GPole} |$}}
\startdata
random & P = 0.4 & P = 0.3\\
$0^\circ - 20^\circ$ & $<0.01$ & $<0.01$\\
$70^\circ - 90^\circ$ & 0.7 & $<0.01$\\
\enddata
\tablecomments{
Probability is a value between 0 and 1 giving the significance level of
the K-S statistics. Small values of probability indicate that the 
CDFs of the two data sets are significantly different from one another.}
\end{deluxetable}

Maser emitting regions are very small in size so that the magnetic fields
traced by masers are not expected to be associated with the
magnetic field in Galactic or molecular cloud scales.
However, there have been some studies showing that the field orientations
measured from the Zeeman effect of OH masers in the star forming regions
have positive correlation with the Galactic scale magnetic field measured from
Faraday rotations \citep[e.g.,][]{Han:2007aa, Fish:2006aa, Noutsos:2012aa}.
Since Class I methanol masers occur in a medium of similar or slightly
less density, and at larger distance from the central stars, than the OH masers,
we have investigated the possibility of association between
the linear polarization of the 44~GHz
methanol masers and the orientation of the Galactic pole, which could
be related to the large scale magnetic field.

We apply the same analysis that is done for the outflows.
As seen in the right panel of Figure~\ref{fig45}, the CDF of the angle
between the polarization angle and the direction of Galactic pole,
$| \chi - PA_{\rm GPole} |$,
is similar neither with the CDF of $0^\circ - 20^\circ$ simulation
nor with that of $70^\circ - 90^\circ$ simulation.
The K-S tests also retune $P < 0.01$ for these two cases (Table~\ref{tb5}),
ruling out the scenario of the polarization angles having preference
on the directions of Galactic plane or pole.

\section{Discussion on Sources with Peculiar Polarization Properties}
We discuss some interesting targets with unique polarization characteristics in detail.

\subsection{G10.32$-$0.26}
\begin{figure}
\plotone{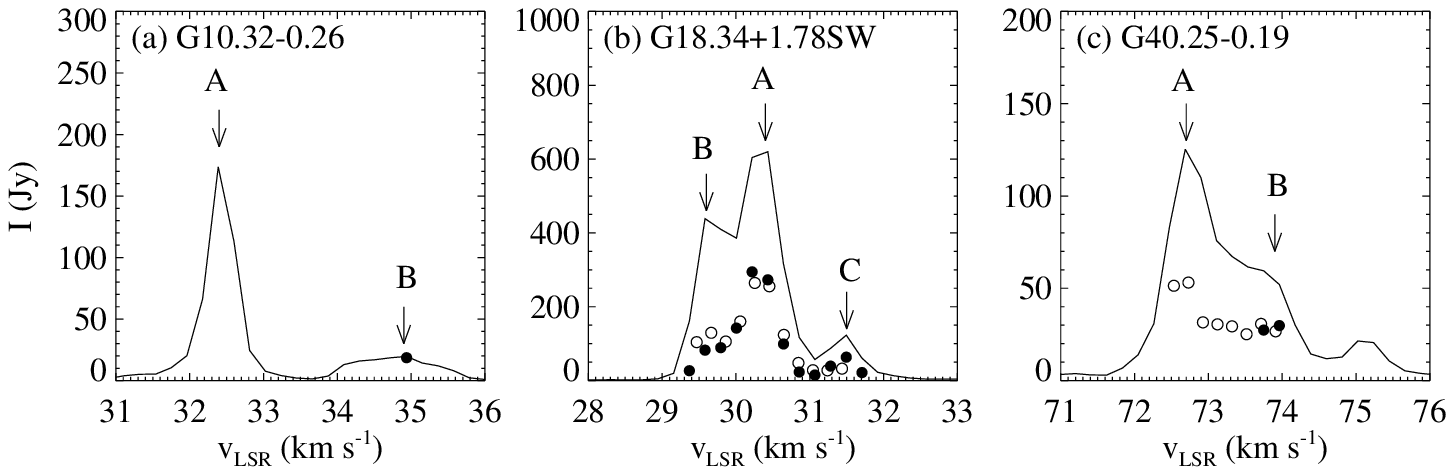}
\caption{Multiple maser components of (a) G10.32$-$0.26, (b) G18.34+1.78SW,
and (c) G40.25$-$0.19 described in Section 5.
The solid line indicates the 44~GHz total flux profile of each source.
Individual components described in the text of Section 5 are marked as
A, B, or C and indicated with arrows.
The filled circles indicate the points with $PI_{44GHz} > 3 \sigma$
after 10 times of multiplication. The open circles are same but at 95~GHz.
\label{fig12}}
\end{figure}

The linear polarization of G10.32$-$0.26 is detected only at 44~GHz
(see Fig.~\ref{fig2}$b$).
It is peculiar because the linear polarization is not detected
in the brightest peak of the total flux profile at \vlsr $\sim 32.5$~\kms
(A component in Fig.~\ref{fig12}$a$)
with an upper limit of 0.9\%, but its secondary peak
at \vlsr $\sim 35$~\kms (B component in Fig.~\ref{fig12}$a$)
shows polarization.
We confirmed this trend by observing them in two or more epochs.
As mentioned in \S~2.1, our observations were targeted for the brightest
features in the total intensity.
If the weaker feature is largely displaced from the brightest one,
for example, by 30\arcsec, that could
produce artificial polarization of up to 2\%.
The displacement of the 35~\kms component from the brightest peak
is less than 1\arcsec, according to the VLBI observation using 
the KVN and VERA combined array (KaVA) (Kim et al. in prep.) 
and its observed degree of polarization is 4 -- 9\%,
which are well above the 2\%. Thus, these detections are very likely real. 

\subsection{G18.34+1.78SW}
G18.34+1.78SW is one of the brightest methanol masers known in the Galaxy
first detected in the KVN single dish maser surveys (Kim et al. in prep.).
It is associated with millimeter core MM2 in a massive star forming region
IRAS 18151-1208 \citep{Marseille:2008aa}.
It is imaged by the KaVA at an angular
resolution of $\sim$2~milli-arcsecond  \citep{Matsumoto:2014aa}.

The single-dish total intensity spectra of G18.34+1.78SW show 3 maser features
with peak velocities of $+30.3$ (A)
$+29.6$ (B), and +31.4~\kms (C in Fig.~\ref{fig12}$b$)
in both transitions,
all of which show linearly polarized emission in both transitions
(see also Fig.~\ref{fig1}$f$).
The velocities of these 3 components are well agreed with the findings
of \citet{Matsumoto:2014aa},
although much of the total flux was missed in the VLBI image.
The polarization properties of these 3 components are summarized in Table~\ref{tb6}.
We used a function with 3 Gaussian components to fit the 0.1~\kms resolution spectra. 
The line widths measured in the polarization profile
range between 0.4 and 0.6~\kms$\!\!$, while it is 0.2~\kms wider when measured in the total flux profile.

This object has $P_L = 2.0 - 6.8$\% in two transitions, and all 3 components
show slightly higher polarization
at 95~GHz, which is generally observed for other sources. 
The polarization angles of the brightest peak are similar at 44 and 95~GHz,
i.e., $12^\circ$ different, but the difference is
beyond the measure error. The angle difference ranges up to
$43^\circ$ in the second brightest peak.
The fact that the polarization angle differences are not consistent
for the 3 maser features in a nearly same position, at most 100 mas
apart \citep{Matsumoto:2014aa}, indicates that neither rotation measure nor the 
beam size can explain the difference in the polarization properties
of the 44/95~GHz transitions. It may be a combination of magnetic field
morphology and the saturation level difference between the two transition
lines as discussed in section 4.2. High resolution line polarimetry
using the VLA or ALMA will help us to understand the underlying physics.

\floattable
\begin{deluxetable}{lcrrrrr}
\tablecaption{Polarization Properties of G18.34+1.78SW\label{tb6}}
\tablehead{
\colhead{$v_{\rm LSR}$} & \colhead{Transition} & \colhead{$I$} & \colhead{$P_L$}&
\colhead{$\chi$} & \colhead{$\Delta v_{\rm FWHM}$}\\
\colhead{( {\rm km s$^{-1}$} )} & \colhead{( GHz )} & \colhead{( Jy )} &
\colhead{( \% )} & \colhead{( $^\circ$ )} &\colhead{( {\rm km s$^{-1}$} )}
}
\startdata
+30.3 & 44&  678. & $4.8\pm0.1 $& $118\pm1$ &   0.5\\
 & 95&  411. & $6.8\pm0.2 $& $130\pm1$ &     0.6\\
+29.6 & 44&  471. & $2.0\pm0.1 $& $96\pm1$ &     0.4\\
 & 95&  346. & $3.6\pm0.2 $& $139\pm1$ &     0.5 \\
+31.4 & 44&  124. & $5.4\pm0.4 $& $117\pm2$  &   0.4\\
 & 95&   50. & $6.2\pm2.4 $& $132\pm8$ &     0.5\\
\enddata
\tablecomments{Spectra with 0.1~\kms resolution were used for Gaussian fit.
The $v_{\rm FWHM}$ presented is the fitting result of polarized intensity
profile, which is 0.2~\kms smaller in maximum than that of
total intensity profile.}
\end{deluxetable}

\subsection{G40.25$-$0.19}
The single dish total intensity profiles of G40.25$-$0.19 show
multiple maser components at \vlsr $= +72 - +75.5$~\kms at both transitions
(see Fig.~\ref{fig1}$l$ and Fig.~\ref{fig12}$c$). This target is interesting because it shows
linear polarization degrees of 4.5\% -- 7\% at 95~GHz in all maser components
at \vlsr $= +72 - +75.5$~\kms$\!\!$, but only a secondary bright feature
at \vlsr $\sim 73.8$~\kms (B in Fig.~\ref{fig12}$c$) is detected at 44~GHz.
The upper limit of the brightest peak at \vlsr $\sim 72.5$~\kms (A)
is 0.8\%.
The difference in polarizaton properties could also be
related to maser saturation level for this source.

\begin{figure}
\plotone{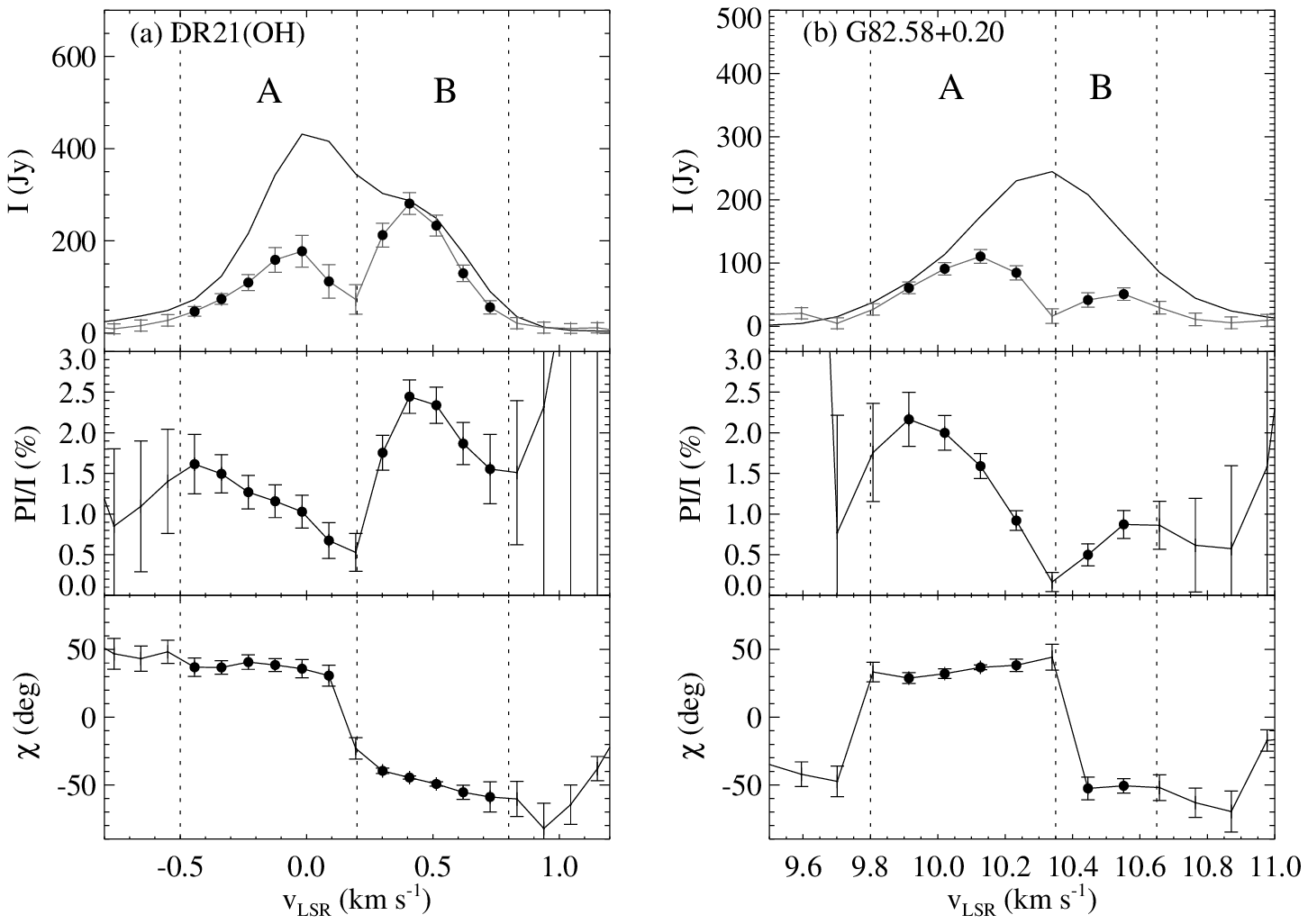}
\caption{
The polarization profiles of
(a) DR21(OH) and (b) G82.58+0.20 at 44~GHz
that show the polarization angle flips as described in Section 5.
The explanations for the figures are same as Fig.~\ref{fig1}, but
zoomed in velocity ranges. The polarized intensities are 40 times multiplied.
Two components showing the $90^\circ$ angle flips are marked as
A and B after divided by dotted lines.
\label{fig13}}
\end{figure}

\subsection{DR21(OH)}
The 44~GHz masers of DR21(OH) have been imaged with the VLA
\citep{Kogan:1998aa, Kurtz:2004aa}. The VLA results show that there are
two groups of masers, one at \vlsr $\sim 0$~\kms and the other
at \vlsr $\sim -4$~\kms$\!\!$, about $20\arcsec$ apart from each other. 
These two groups of masers are located at the end of approaching and
receding outflow in the DR21(OH) region~\citep{Zhang:2014aa},
supporting the theory of Class I methanol masers being stimulated by outflows.
Our observation was pointed to the brightest peak at \vlsr $\sim 0$~\kms
that contains $\sim 80$\% of the total flux and
is located at the end of an receding outflow lobe.
This feature has an elongated, arc-shape morphology in the VLA map.

The linear polarization of DR21(OH) shows features indicating depolarization
and a sudden flip of polarization angle in the 44~GHz transition line.
The total intensity and the polarized emission of DR21(OH) at 44~GHz show
two maser components, one at \vlsr = $-0.5 - +0.2$~\kms and the other
at \vlsr = +0.2 -- +0.8~\kms, which are indicated as A and B regions
in the panel (a) of Figure~\ref{fig13}.
Their polarization fractions are similar to be $\sim 2$\%, while their
polarization angles are different by $\sim 90^\circ $.
At the velocity where the two components overlapped, the fraction
of linear polarization drops to minimum.

Taking into account the previous observations,
elongated masers aligned along a curved receding shock front
with the magnetic field compressed along the shock can explain the
observed profiles.
Such flip of polarization angle is expected around $\theta = 55^\circ$
in maser theories\citep[e.g.,][]{Goldreich:1973aa},
and has been observed previously in SiO masers
by \citet{Kemball:1997aa} and in H$_2$O by \citet{Vlemmings:2006aa}.
In these observations, the angle flip appeared in a single maser feature,
which was interpreted as an evidence of curved magnetic field with $\theta$
being close to the van Vleck angle, i.e., $\theta = 55^\circ$.
In our case, the flip happens in two different maser features
separated in velocities in a single dish spectrum,
but the underlying physics could be the same.
If the two maser features are located in a curved magnetic field aligned along
the shock front receding from us, each of their $\theta$ being
$\theta < 55^\circ$ and $\theta >55^\circ$, then, they would
produce a 90$^\circ$ angle flip as well as a LOS velocity shift
that are observed.
One implication of this explanation is that the 44~GHz masers of
DR21(OH) is unsaturated or only moderately saturated
according to the simulation of \citet{Nedoluha:1990aa}.
In the upper panels of Figure 3 in their paper,
the polarization angle ($\phi$ in their paper) flips by the change of $\theta$
only when $\log R / \Gamma \lesssim 1$. When the maser is highly saturated,
e.g., when $log R / \Gamma = 3$, the polarization angle is not seriously
affected by $\theta$ but rather constant.

The polarization angle can also be different as large as 90$^\circ$ when
the saturation levels of the two maser components are different.
As seen in the Figure 3 of above paper, for example,
if the saturation levels of the two components are $\log R / \Gamma$ = 1 and 3,
respectively, the two components would show 90$^\circ$ of polarization angle
difference even for the same field orientation with $\theta = 15^\circ$.
How much the saturation levels of the same 44~GHz transition line
can be different is another issue. Assuming the not-well-known factor
$\Gamma$ being similar, $R$ would be determined by the
physical size and the flux of maser components.
The difference of the observed fluxes are less than factor of two different,
while their physical sizes cannot be determined with current observations.

The angle flip is not observed in the 95~GHz transition line profile
(see Fig~\ref{fig1}$n$).
While the polarization angle of the $-0.5 - +0.1$~\kms component at 95~GHz
is almost perpendicular to that of the 44~GHz maser, the angle of the
$+0.3 - +0.7$~\kms component is similar to that of the 44~GHz maser.
If the angle flip in the 44~GHz maser is due to the van Vleck angle crossing,
the polarization angles of the 95~GHz maser being relatively constant
may imply that the 95~GHz transition is more highly saturated than
the 44~GHz transition, in which condition, the position angle is always
perpendicular to the magnetic field regardless of $\theta$,
as \citet{Nedoluha:1990aa} demonstrated.
The saturation levels of maser are difficult to be measured. 
Future high resolution observations and the theoretical study dedicated to the
methanol maser lines would be able to reveal the morphology, field orientation,
and kinematics of the DR21(OH) region as well as the involved maser physics.

\subsection{G82.58+0.20}

The polarization profiles of G82.58+0.20 present polarization intensity
and angle variation similar to those of DR21(OH) at both transition lines.
It shows polarization angle flip and the reduced polarization fraction
at velocities where the two velocity components overlapped
in the 44~GHz transition line,
as indicated as A and B regions in the panel (b) of Figure~\ref{fig13}.
This may imply the existence of elongated masers associated
with curved magnetic field orientation in a compressed medium.
This target would be a good candidate for the high resolution
polarimetry observations. 

\section{Summary}
We performed the first comprehensive study of the linear polarization
of a large sample of 39 bright
Class I methanol maser sources in the 44 and 95~GHz transitions simultaneously.
The main findings of this study are summarized as follows. 

\begin{enumerate} 
\item 
We detected 23 sources (59\%) in at least one transition and 15 sources
(38\%) in both transition. Their error-weighted mean degrees of polarizations
are $2.7\pm 0.3$\% and $4.8\pm 0.1$\% at
44 and 95~GHz, respectively. The linear polarization of the 44~GHz methanol
maser transition was first detected in this study.
\item
We compared the polarization properties of the 44 and 95~GHz transition lines.
We found that their polarization fractions are linearly correlated,
but the emission from the 95~GHz tends to be more linearly polarized than
that from the 44~GHz transition line. We suggest that this is not likely
due to the external reason, such as Faraday rotation or different beam
size effect. The remaining possibility is
the intrinsic differences of the physical parameters
involved in the maser process, such as the decay rate, 
Zeeman rate, maser size, and beaming angle.
\item
The polarization angles of the 44 and 95~GHz transition lines
are well correlated in general, implying that both transitions
are experiencing similar magnetic environment.
\item
We did not observe any source with a fractional polarization $> 30$\%.
Such high fractions were found by \citet{Wiesemeyer:2004aa} and
require anisotropic pumping/loss conditions.
Our observations show that such conditions may be not as common in the
star forming regions as \citet{Wiesemeyer:2004aa} suggested.
\item
We found that the polarization angles tend to be aligned perpendicular
to the outflows for the 7 maser sources with known outflow orientations.
This requires more samples to be generalized.
We also found that the maser polarization angles are not particularly
correlated with the Galactic geometry.
\item
We discussed some targets with peculiar polarization properties.
Among them, DR21(OH) and G82.58+0.20 appear to be interesting targets
because they show the $90^\circ$ polarization angle flip in the 44~GHz
polarization profiles,
while it is not visible in their 95~GHz polarization profiles.
The high angular resolution polarimetry observations in both frequencies and
the theoretical studies will reveal whether these polarization properties
are due to the van Vleck angle crossing or change of maser saturation level,
providing more information on the not-well-known methanol maser polarization
physics.
\end{enumerate}

\acknowledgements
We are grateful to all staff members in KVN who helped to operate the array and to correlate the data. The KVN is a facility operated by KASI (Korea Astronomy and Space Science Institute). The KVN operations are supported by KREONET (Korea Research Environment Open NETwork) which is managed and operated by KISTI (Korea Institute of Science and Technology Information).
W.V. acknowledges support from ERC consolidator grant 614264.
This research has made use of the SIMBAD database,
operated at CDS, Strasbourg, France. 



\end{document}